\documentclass[journal=jacsat,manuscript=article, layout = onecolumn]{achemso}
\usepackage[version=3]{mhchem} 
\usepackage{gensymb}
\usepackage{rotating}
\usepackage{booktabs}
\usepackage{multirow} 
\usepackage{makecell} 
\usepackage{chemformula} 
\usepackage{capt-of} 
\usepackage{textcomp}
\usepackage{siunitx}
\usepackage{xr}
\usepackage{amsmath}
\usepackage{amssymb}

\usepackage[colorlinks, urlcolor=blue, citecolor=blue]{hyperref}
\hypersetup{
    colorlinks=true,
    linkcolor=blue,
    filecolor=blue,      
    urlcolor=blue,
    }

\author{Çayan Demirkır}
\affiliation[University of Twente]
{Physics of Fluids, University of Twente, Enschede  7500 AE, The Netherlands}

\author{Jeffery A. Wood}
\affiliation[University of Twente]
{Soft Matter, Fluidics, and Interfaces, University of Twente, Enschede  7500 AE, The Netherlands}

\author{Detlef Lohse}
\affiliation[University of Twente]
{Physics of Fluids, University of Twente, Enschede  7500 AE, The Netherlands}
\alsoaffiliation[Max Planck Institute for Dynamics and Self-Organization]
{Max Planck Institute for Dynamics and Self-Organization, Am Fassberg 17, 37077 Göttingen, Germany}

\author{Dominik Krug}
\email{d.j.krug@utwente.nl}
\affiliation[University of Twente]
{Physics of Fluids, University of Twente, Enschede  7500 AE, The Netherlands}

\title{\vspace{-2.3cm}Life beyond Fritz: On the detachment of electrolytic bubbles}

\abbreviations{IR,NMR,UV}

\begin{document}

\begin{abstract}
We present an experimental study on detachment characteristics of hydrogen bubbles during electrolysis. Using a transparent (Pt or Ni) electrode enables us to directly observe the bubble contact line and bubble size. Based on these quantities we determine other parameters such as the contact angle and volume through solutions of the Young-Laplace equation. We observe bubbles without ('pinned bubbles') and with ('spreading bubbles') contact line spreading, and find that the latter mode becomes more prevalent if the concentration of $\ce{HClO4}$ $\geq 0.1~\mathrm{M}$. The departure radius for spreading bubbles is found to drastically exceed the value predicted by the well-known formula of W. Fritz (\textit{Physik. Zeitschr.} \textbf{1935}, 36, 379–384) for this case. We show that this is related to the contact line hysteresis, which leads to pinning of the contact line after an initial spreading phase at the receding contact angle. The departure mode is then similar to a pinned bubble and occurs once the contact angle reaches the advancing contact angle of the surface. A prediction for the departure radius based on these findings is found to be consistent with the experimental data. 
\end{abstract}

\section{INTRODUCTION}
\subsection{Motivation}
Hydrogen possesses great versatility as a clean energy carrier and holds the potential to substantially curtail carbon emissions across diverse sectors, including transportation and industrial operations \cite{Rosen_2016, MacFarlane2020}. Electrolysis is one of the most promising hydrogen production methods, utilizing renewable electricity sources while avoiding greenhouse gas emissions. Nevertheless, the cost of producing clean hydrogen remains notably higher compared to fossil fuels \cite{Shiva_Kumar_2022,Swiegers2021}. Several factors contribute to considerable energy losses and low efficiency during the electrolysis process \cite{Brauns_2020}. Notably, the presence of gas bubbles on the electrode surface significantly diminishes the active area and creates undesirable resistance, resulting in bubble overpotentials \cite{Zhao_2019, Raman_2022, Vogt_1983, Vogt_2012}. Studies indicate that removing these bubbles can significantly decrease the required applied potential on the electrolyzer \cite{Angulo_2020, Zeng_2010, Swiegers2021}. Hence, comprehending the dynamics of electrolytic bubbles becomes crucial in developing novel techniques to enhance the efficiency of the electrolysis process.

The lifetime of an electrolytic bubble has been investigated in many studies \cite{Raman_2022, Zhao_2019, Brandon_1985, Sequeira_2013, Lubetkin_1995, Yang_2015}, and is mainly divided into the following stages: nucleation, growth, and detachment. Typically, the nucleation of a bubble spontaneously occurs on a cavity or surface inhomogeneity once the supersaturation of the gases near the electrode reaches a critical threshold  \cite{Borkent_2009, Jones_1999, German2018}. Following nucleation, the bubble grows on the electrode surface. The growth rate is determined by factors such as pressure, diffusion rate or surface reactions \cite{Westerheide1961, Brandon_1985, van_der_Linde_2017, Bashkatov_2022, Khalighi2023, Raman_2022}. The three-phase (gas-liquid-solid) contact line either remains pinned at the cavity edge throughout the bubble lifetime (`pinned' or `cavity' bubble) or eventually starts spreading over the electrode surface (spreading bubble) while the bubble is growing \cite{Chesters_1977}. Finally, the detachment takes place through either the coalescence of multiple bubbles or the individual detachment of a single (or isolated) bubble due to the buoyancy \cite{Iwata_2022}. The size of the three-phase contact line shows a considerable difference between pinned and spreading bubbles. This has a direct influence on the detachment radius ($R_{det}$) because the contact line, along with the contact angle ($\theta$), determines the adhesion force that keeps bubble attached the electrode \cite{Antonini2009}. Therefore, understanding the contact line dynamics of a bubble is substantial, however the detection of the contact line and tracking its development on an electrode is not straightforward. For a single bubble that grows on a micro/nano structured electrode, the size of the contact line radius ($R_{cont}$) is intrinsically limited by the pit radius \cite{Raman_2022, van_der_Linde_2017, Penas2019}. On the other hand, for a planar electrode, many bubbles grow together during the electrolysis, and it can be challenging to optically detect the contact line radius ($R_{cont}$) of a bubble due to blockage by other bubbles. Hence, monitoring $R_{cont}$ at a planar electrode requires a transparent electrode configuration that allows the bubble's contact line to be seen from the back of the electrode. There are few examples of a similar configuration in the literature \cite{Fernandez_2014, Matsushima_2013, Janssen_1981, Janssen_1984, Pande_2020, Pande_2021, Sides_1985}, but none of them actually focused on the contact line dynamics of the bubbles. 

Extensive research in the literature has explored various factors influencing the detachment event for both coalescence-driven and buoyancy-driven cases (e.g., electric potential, electrolyte type and concentration, external flow, surfactants, electrode morphology, and surface wettability, etc.)\cite{Bashkatov_2022, Matsushima_2006, Janssen_1973, Brandon_1985, Park_2023, Abdelghani_Idrissi_2021, Zhang_2012, Fernandez_2014, Ahn_2013, Allred_2021, Hong_2011, Kim_2017, Sepahi2024}. However, the prediction of the detachment radius ($R_{det}$) by unraveling the mechanisms of the more fundamental case, the detachment of an isolated bubble from a horizontal planar electrode with no external flow holds a significant importance in advancing the strategies for electrolytic bubble removal. 

In this study, we investigated the detachment characteristics of the hydrogen bubbles formed in water electrolysis, focusing on the contact line dynamics and considering the effects of the current density, electrolyte concentration (or pH), and electrode material. Experiments were carried out across a range of concentrations from $10^{-4}$ M to 1 M $\ce{HClO4}$, with nominal current densities ranging from 10 to 200 $\ce{A/m^2}$ and utilizing platinum and nickel as electrode materials. First, the factors influencing the contact line formation (pinned or spreading) are investigated. In addition, we reveal the dynamics of the contact line during the growth of a spreading bubble, and discuss the effects of the dynamic contact angles on the contact patch size, bubble morphology and detachment radius. 

\subsection{Theoretical Background}

\subsubsection{Bubble shape and contact angle}

 \begin{figure}[ht]
		\centering
		\includegraphics[width=0.5\columnwidth]{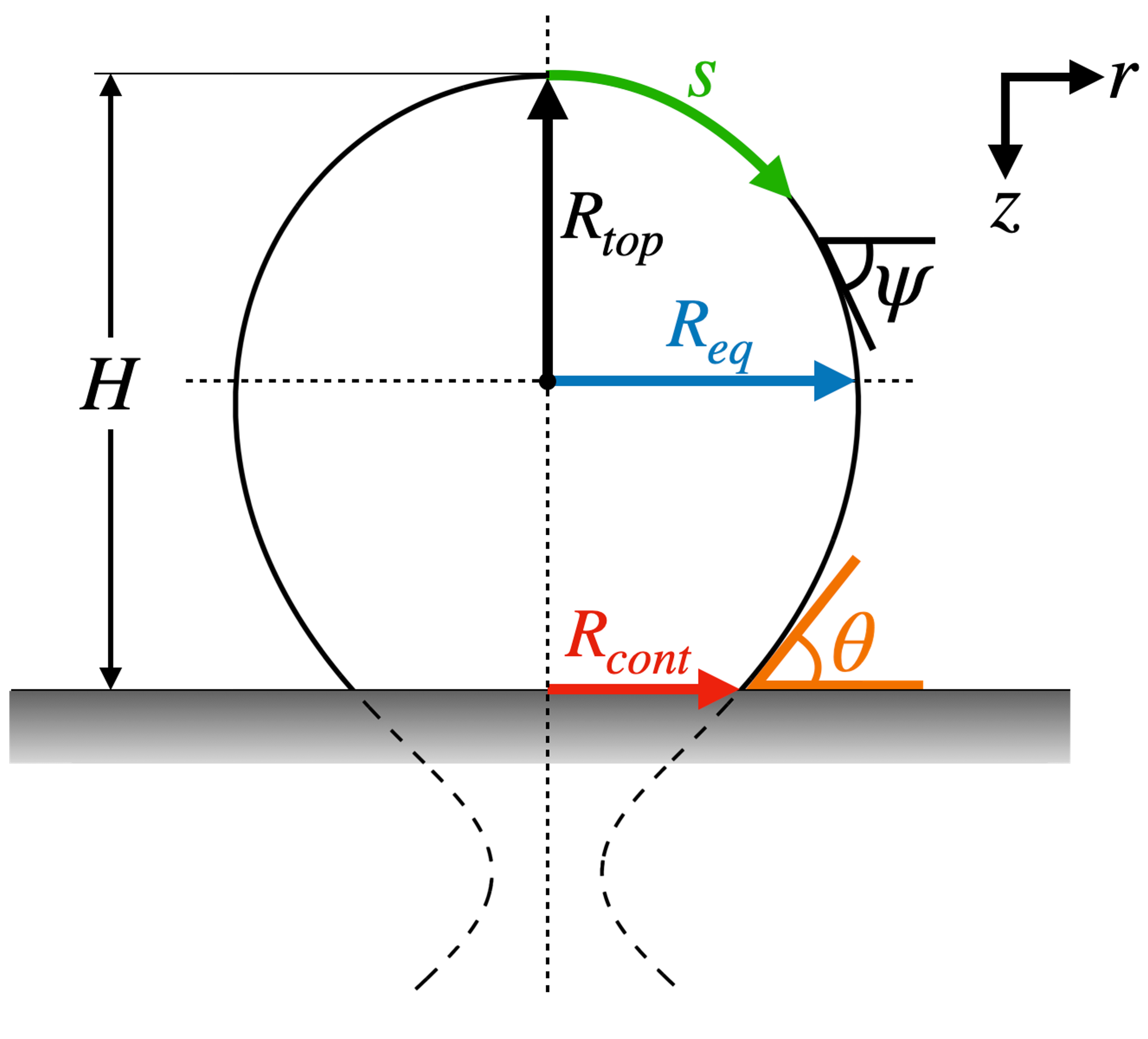}
		\caption{Example of a bubble shape obtained from the Young-Laplace (YL) equation (\ref{eq:YL1}) at $Bo =  0.3$.}
		\label{fig:YL_shape}
	\end{figure}

When considering only the effects of buoyancy and capillarity (i.e. neglecting potential additional contributions from electric or Marangoni forces), the shape of an equilibrium bubble is described by the Young-Laplace (YL) equation\cite{Bashforth_1883}
    \begin{equation}
    \frac{d\psi}{d\bar{s}} = 2 - Bo\cdot\bar{z} -\frac{\sin{\psi}}{\bar{r}}.
    	\label{eq:YL1}
    \end{equation}
Here,
      \begin{equation}
          Bo = \frac{\Delta \rho g R_{top}^2}{\sigma} 
      \end{equation}
is the Bond number.  $\Delta \rho = \rho_l - \rho_g \approx \rho_l$ denotes the density difference between the liquid ($\rho_l$) and the gas ($\rho_g)$ density, and $\sigma$ is the surface tension. The coordinates $z$ and $r$ point along the centerline and the radial direction, respectively with the origin fixed at the bubble top as shown in \hyperref[{fig:YL_shape}]{Figure \ref{fig:YL_shape}}. All length scales are normalised by the curvature $R_{top}$ at the apex indicated an overline, e.g. $\mathrm{\bar{z} = z/R_{top}}$. The coordinates are geometrically related via
    \begin{equation}
        \frac{d\bar{r}}{d\bar{s}} = \cos\psi \quad \textrm{and } \quad \frac{d\bar{z}}{d\bar{s}} = \sin\psi,
      	\label{eq:YL2}
      \end{equation}
      where $s$ denotes the arc length. The system of equations (\ref{eq:YL1}) and (\ref{eq:YL2}) can be solved numerically to yield the bubble shape as a function of the Bond number. The boundary condition at the surface is applied by ending the solution at the appropriate contact angle ($\theta$) or contact patch radius ($R_{cont}$), as shown in \hyperref[{fig:YL_shape}]{Figure \ref{fig:YL_shape}}.

\subsubsection{Force balance}
Under the same assumptions leading to equation (\ref{eq:YL1}), the forces in the $z$-direction for a bubble on a horizontal surface are given by (see the SI for derivation and sketch of the forces) 
\begin{align}
F_{b} &= -V \cdot (\rho_l - \rho_g)\cdot g,
    \label{eq:buo}\\
    F_{corr} &= -\pi R_{cont}^2\left(\frac{2\sigma}{R_{top}}-\rho_l gH\right),
    \label{eq:corr}\\
    F_{s} &= 2\pi R_{cont}\sigma {\sin}{\theta}.
    \label{eq:surf_tens}
\end{align}
Here, $V$ denotes the bubble volume, $H$ the bubble height about the surface as shown in \hyperref[{fig:YL_shape}]{Figure \ref{fig:YL_shape}}, and $g$ is the gravitational acceleration. Out of the three forces, both buoyancy ($F_b$) and the pressure correction force ($F_{corr}$) point upwards (negative $z$ direction) and are therefore defined negative here, while the surface tension force ($F_s$) acts to retain the bubble. Since the bubble is and remains at rest 
\begin{equation}
\sum{F_z} = F_b + F_{corr} + F_s  = 0
     \label{eq:force_bal}
\end{equation}
at \emph{all times} while the bubble is sitting on the electrode. The force balance alone is therefore not sufficient to predict bubble departure and must be supplemented by a suitable departure criterion, which depends on the contact line dynamics. 

\subsubsection{Pinned bubbles}
We will refer to cases where the contact line remains pinned, e.g. to the edge of the cavity at which it nucleated. In this case, the shape is given by solutions of equation (\ref{eq:YL1}) with $R_{cont} = \textrm{const.}$, and the relevant detachment criterion is for the contact angle to reach a value of $90^\circ$\cite{Chesters_1977}. Since typically the volume equivalent bubble radius ($R_b$) significantly exceeds $R_{cont }$ at this point, $F_{corr}$ can be neglected and the force balance is between $F_b$ and $F_s$ only. This is confirmed by \hyperref[{fig:cav_fritz_force}]{Figure \ref{fig:cav_fritz_force}}a, where the force balance for a pinned bubble with $R_{cont} = \SI{2}{\micro\meter}$ is shown until departure. Initially when ($R_{cont } \approx R_{top}$), $F_{corr}$, the bubble volume ($V$) and hence $F_{b}$ are small and the balance is between $F_s$ and $F_{corr}$. The pressure force decreases as the Laplace pressure in the growing bubble reduces and buoyancy becomes the dominant upward force on the bubble. The surface tension force $F_s$ initially decreases as the contact angle decreases from its initial value of $\theta =90^\circ$ when $R_{top}= R_{cont}$. At later stages of the bubble growth, $\theta$  and therefore also $F_s$ increase again until departure occurs when  $\theta =90^\circ$ is reached again. Evaluating equation (\ref{eq:force_bal}) for the balance between $F_b$ and $F_s$ at departure, it then follows that 
\begin{equation} \label{eq:cav_det}
			R_{cav} = \left(\frac{3}{2}\frac{\sigma R_{cont}}{\rho g}  \right) ^ \frac{1}{3},
\end{equation}
where $R_{cav}$ is the volume equivalent sphere radius to the volume $V_{max}$ of the pinned bubble with $\theta = 90^\circ$. Note that we use $V_{max}$ to denote the maximum volume of the bubble before departure for either pinned or spreading bubbles.

\subsubsection{Spreading bubbles}
\begin{figure}[ht]
		\centering
		\includegraphics[width=1\columnwidth]{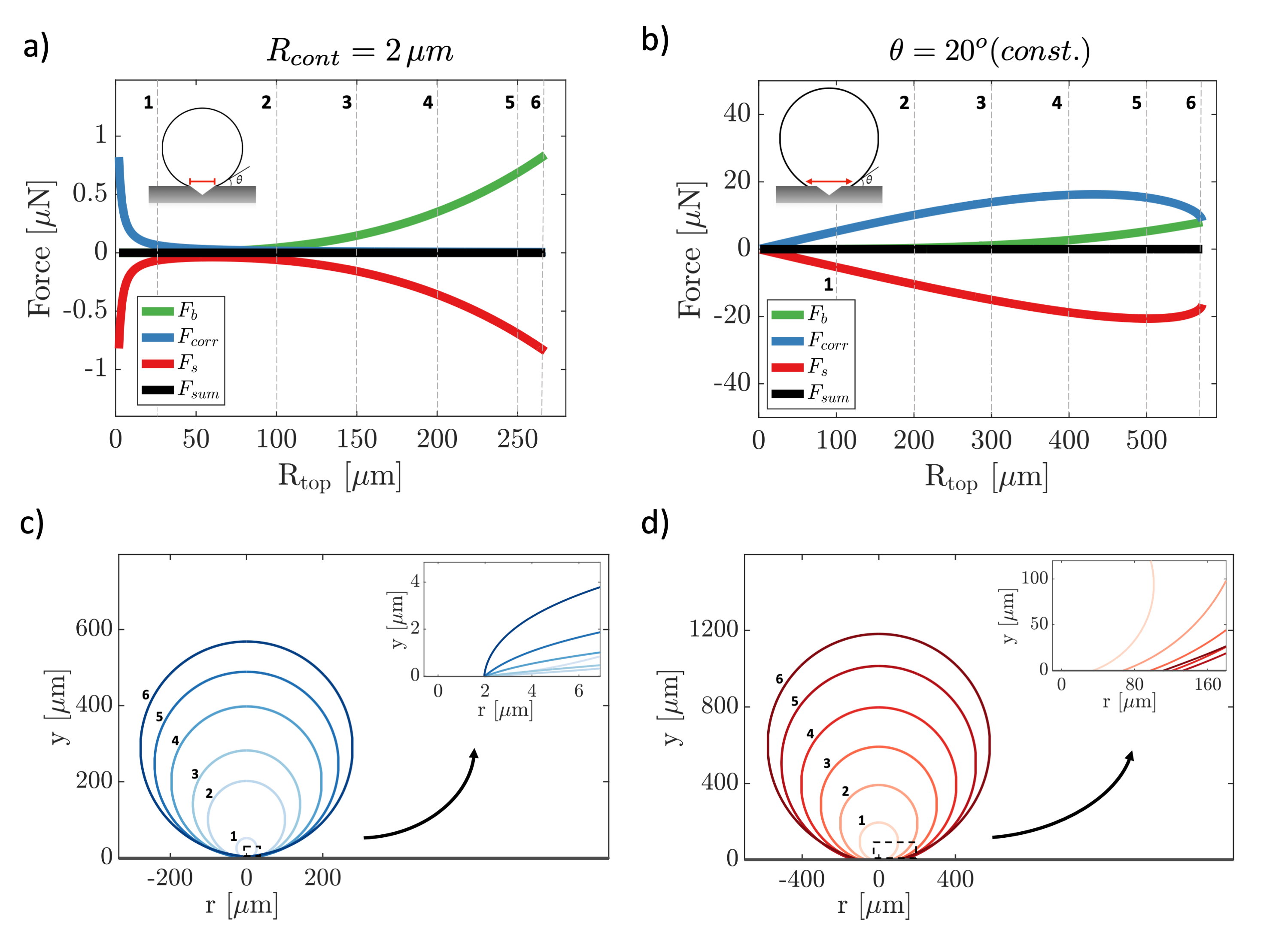}
		\caption{The concept sketches (insets) and the force evolution for pinned (a) and spreading (b) bubbles. The forces acting on the pinned bubble were computed for a contact radius ($R_{cont}$) of \SI{2}{\micro\meter}, while those for the spreading bubble were computed assuming a constant contact angle ({$\theta$}) of 20$\degree$. The corresponding bubble shapes (labelled by the numbers) including a zoom in on the foot region in the insets are shown in panels (c) (pinned case) and (d) (spreading case).}
		\label{fig:cav_fritz_force}
\end{figure}

If a bubble grows from a cavity, the contact angle initially decreases. For the minimum contact angle ($\theta_{min}$), \citet{Chesters_1978} obtained $\sin \theta_{min}= (4/3)\cdot 2^{1/4}\sqrt{R_{cont}/\lambda_c}$, where $\lambda_c = \sqrt{\sigma/(\Delta \rho g)}$ is the capillary length. It was argued that in the absence of contact angle hysteresis, 
the bubble starts to spread on the surface with constant $\theta$, if $\theta_{min}$ is lower than the static (or equilibrium) contact angle of the surface.  In this case, it can be shown that $V_{max}$ is reached when the inflection of the profile 
determined from equation (\ref{eq:YL1}) is located at the interface \cite{Chesters_1978}. Even for a low contact angle of $\theta = 20^\circ$, the pressure correction force remains relevant and even exceeds $F_b$ as \hyperref[{fig:cav_fritz_force}]{Figure \ref{fig:cav_fritz_force}}b shows. In 1935, \citet{Fritz_1935} determined $V_{max}$ for spreading bubbles as the maximum volume for which a solution of equation (\ref{eq:YL1}) for a given value of $\theta$ exists. He based his findings on the numerical solutions of \citet{Bashforth_1883} and determined the famous linear relation
\begin{equation} \label{eq:Fritz}
			R_{Fritz} = 0.0104  {\theta}  \lambda_c,
\end{equation}
between the volume-equivalent detachment radius ($R_{Fritz}$) and $\theta$ given in degrees.

\section{MATERIALS AND METHODS}
\subsection{Electrochemical Cell}

\begin{figure}[ht]
		\centering
		\includegraphics[width=1\columnwidth]{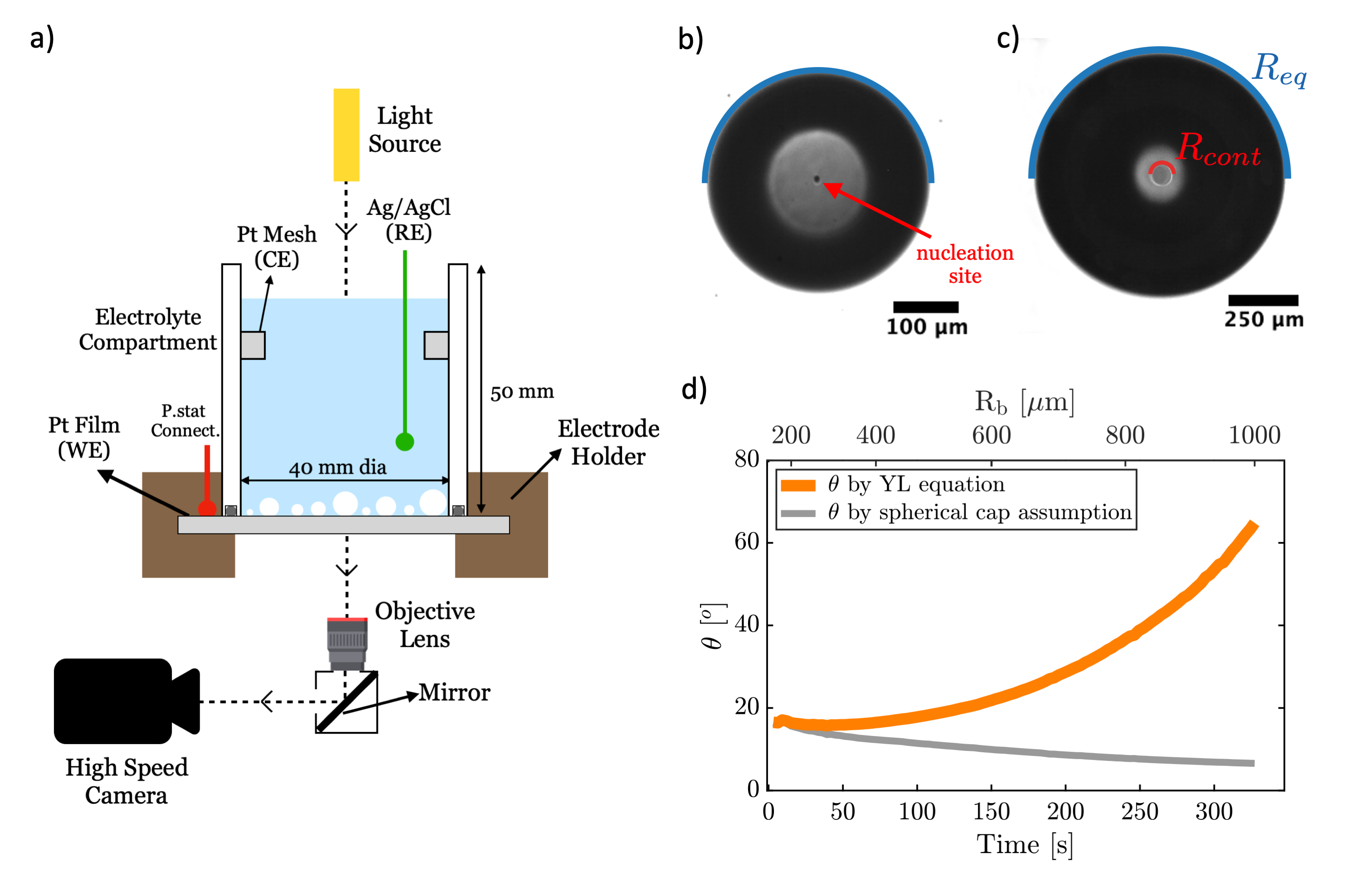}
		\caption{The sketch of the experimental setup (a), typical experimental images of pinned (b) and spreading (c) bubbles, and development of the contact angle ({$\theta$}) for a spreading bubble obtained by different methods (d) are shown as function of time (bottom) and bubble size (top axis). The red half circle in (c) shows the contact area, and blue half circles on (b, c) represent the bubble size. The nucleation site of the pinned bubble is shown by an arrow. The bright areas with diffuse edges in the centre of the bubbles are caused by light passing through the bubble and reaching the camera. }
		\label{fig:setup_exp_images}
	\end{figure}

A lab scale electrochemical cell with a disk working electrode (WE) was built for the electrolysis experiments (\hyperref[{fig:setup_exp_images}]{Figure \ref{fig:setup_exp_images}}a). The cylindrical electrolyte compartment made of Teflon (PTFE) has an inner diameter of 40 mm and a height of 50 mm. An O-ring is placed between the WE and the bottom wall of the Teflon component, and the cell is squeezed by another part that is made of PEEK (polyetheretherketone) to ensure the sealing. All the materials in contact with the electrolyte were selected to be compatible with the acid solutions. Similar to earlier work \cite{Pande_2020,Pande_2021}, the WE was produced by sputtering a 20 nm thin film of either platinum or nickel onto a glass slide. The connection between WE and potentiostat was ensured by using a point contact. Unless otherwise stated, platinum was used as the main electrode material in this study. However, some experiments were also performed using a nickel electrode to demonstrate the material-independent behaviour of the contact line behaviour of the bubbles. For better adhesion, a 3 nm tantalum layer was applied between the film and the glass slide. The thickness of 20 nm proved to provide the best compromise, allowing for sufficient transparency to observe the contact line dynamics while keeping the sheet resistance low.  The sheet resistance of the electrode used in present study was measured to be $\approx$ 50 $\Omega$ using a multimeter. This value is consistent with previous work \cite{Pande_2020}, where the sheet resistance of a 10 nm thick disk electrode was measured to be 69 $\Omega$ using the four probe method.

A platinized titanium mesh with a significantly larger surface area than the WE was utilized as the counter electrode (CE), and positioned approximately 3.5 cm above the WE. An Ag/AgCl electrode (in 3 M NaCl; BASi{\textsuperscript{\tiny\textregistered}}) was placed close ($\approx$ 5 mm) from the WE and served as the reference electrode (RE). Perchloric acid ($\ce{HClO4}$) solutions were prepared in different concentrations from $10^{-4}$ M to 1 M, corresponding to a pH range from about 4 to 0, respectively. To increase the electrical conductivity of the electrolyte, sodium perchlorate monohydrate ($\ce{NaClO4.H2O}$) salt was added to achieve a supporting electrolyte concentration of 0.5 M (except in the 1 M $\ce{HClO4}$ case). Since the concentration of the 1 M $\ce{HClO4}$ electrolyte was sufficiently high, no supporting electrolyte was added.  The chemicals were supplied from Sigma-Aldrich (purity of 99.99\%).
 
\subsection{Methods}

The experiments were done under ambient pressure, at constant current densities from 10 to 200  $\ce{A/m^2}$ using a Biologic VSP-300 potentiostat. In order to minimize the dissolved $\ce{O2}$ in the electrolyte, $\ce{N2}$ purging was applied for at least 30 minutes before every experiment. During electrolysis, bubble images were recorded by a high-speed camera (Photron Nova S12) with frame rates between 10 Hz and 15,000 Hz, depending on the dynamics we were interested in. The camera was positioned horizontally, and the optical path was redirected vertically through the WE by a mirror at an angle of 45\degree. The resolution was chosen to resolve as much detail of the contact line dynamics as possible while also capturing the bubble outline. 

For the smaller bubbles at electrolyte concentrations up to $10^{-2}$ M, we employed a 10x magnification objective lens (Olympus LMPLFLN) yielding a resolution of \SI{1.78}{\micro\meter}/px. At higher concentrations, we switched to a 5x magnification objective lens (Olympus MPLFLN, \SI{3.52}{\micro\meter}/px) to accommodate the larger bubbles within the field of view. Backlight illumination was provided by a  light source (Schott KL2500). A schematic of the entire optical arrangement is shown in \hyperref[{fig:setup_exp_images}]{Figure \ref{fig:setup_exp_images}}a. The calibration of both optical systems was performed by imaging beads with diameters of $\sim$1 mm and $\sim$2 mm placed in the electrolyte compartment filled with water. 15 beads for each size were randomly chosen, and their sizes were determined  via a calibrated DSLR camera (Nikon D850). The mean diameters were found as 1.017 mm and 1.994 mm with standard deviations of \SI{1.4}{\micro\meter} and \SI{4.8}{\micro\meter}, respectively.  Given that the size of the bubbles was considerably smaller than the diameter of the disk electrode (40 mm), multiple bubbles were formed on the electrode surface once the experiment was started. Since the field of view only covered a small part of the electrode area, we searched for isolated bubbles (no coalescence with another bubble) by moving the lateral position of the cell with a micro stage on which we placed the cell. The focal plane was carefully adjusted in the vertical plane using a motorised stage to be on the electrode surface (minimum increment is \SI{0.05}{\micro\meter}).

In \hyperref[{fig:setup_exp_images}]{Figure \ref{fig:setup_exp_images}}(b, c), the typical experimental images of pinned and spreading bubbles, respectively, are shown. In both instances, the blue curve indicates the outline of the bubble shadow, which corresponds to the equator of the bubble (see \hyperref[{fig:YL_shape}]{Figure \ref{fig:YL_shape}}) and we denote the corresponding radius as $R_{eq}$. Due to the bubble curvature, only a portion of the light can reach to the camera, resulting in a bright area with a diffuse edge at the center, as seen in both images. For the pinned bubble in \hyperref[{fig:setup_exp_images}]{Figure \ref{fig:setup_exp_images}}b, a small nucleation site (indicated by the arrow) can be identified but no contact line is visible during the entire evolution of this bubble. In contrast, a clear contact line is visible for the spreading bubble in \hyperref[{fig:setup_exp_images}]{Figure \ref{fig:setup_exp_images}}c, based on which we can experimentally determine $R_{cont}$. 

Image processing was carried out by the image processing toolbox of MATLAB, following these steps: Firstly, the region of interest was cropped from the original image to encompass the entire size of a bubble on that particular image. To avoid the cases where optical path obstruction from other detached bubbles occurred and edge detection becomes challenging, the contrast of the image was adjusted by specifying the intensity limits. Then, binarization was performed twice separately for the contact line area and bubble size in order to increase the sensitivity of the detection. The quality of the binarization was checked by visual inspection of the binarized image overlaying the original image. 

We utilise the YL equation to obtain other parameters such as the bubble volume (V) and the contact angle ($\theta$) based on the experimentally obtained values of $R_{eq}$ and $R_{cont}$. To this end, we initially assume that $R_{top} \approx R_{eq}$, i.e. that the bubble shape is approximately spherical at the top, to obtain an estimate of $Bo$, based on which equation (\ref{eq:YL1}) is solved for a first estimate of the bubble shape. Based on the difference between the measured and modelled values of $R_{eq}$ the Bond number is adjusted slightly until the two match. The contact angle is then determined from the inclination of the curve at the point where the radius equals $R_{cont}$ (see \hyperref[{fig:YL_shape}]{Figure \ref{fig:YL_shape}}). As shown in \hyperref[{fig:setup_exp_images}]{Figure \ref{fig:setup_exp_images}}d, the results from this procedure drastically deviate from the simpler (but incorrect) alternative\cite{Yang_2015, Fernandez_2014} in which simply a spherical shape of the bubble is assumed. In that case, the contact angle is given by $\theta = \sin^{-1}(\mathrm{{R_{cont}}/{R_{eq}}})$. However, the spherical cap assumption is inconsistent with the force balance in equation (\ref{eq:force_bal}), since it leads to $F_s + F_{corr} \approx 0$. The present method is therefore more appropriate.

To characterise the surface properties, contact angle and surface tension measurements were performed using sessile and pendant drop techniques, respectively. Additionally, atomic force microscopy (AFM) and scanning electron microscopy (SEM) were employed to characterize the surface conditions of the electrodes both before and after use. Details and additional results from these measurements are included in the Supporting Information.

\section{RESULTS AND DISCUSSION}

\subsection{Factors Affecting Contact Line Dynamics}

\begin{figure}[ht]		\centering
		\includegraphics[width=1\columnwidth]{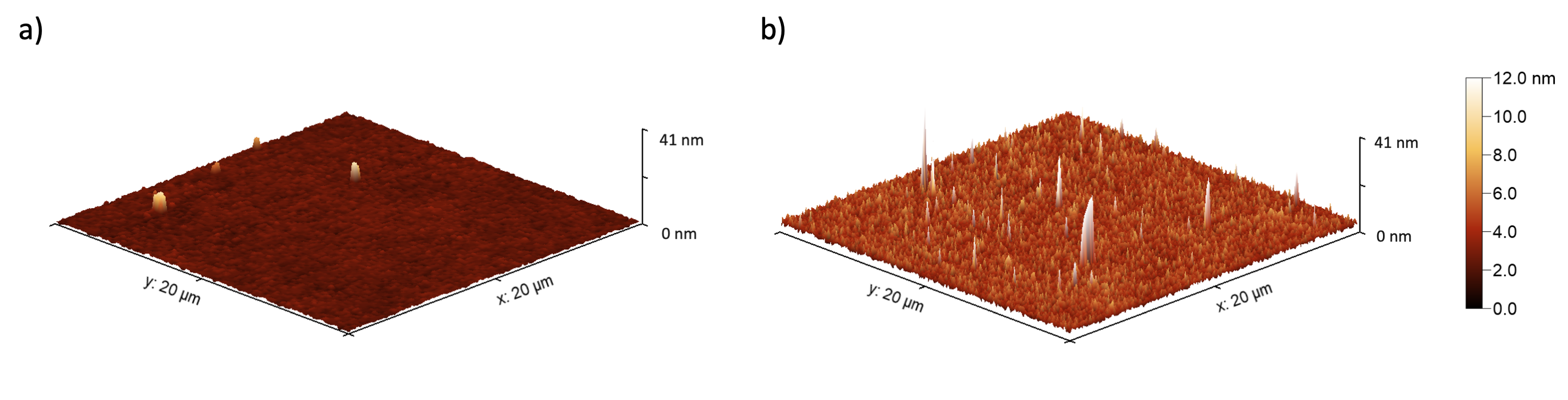}
		\caption{Atomic forced microscopy (AFM) measurements of a new (a) and used (b) electrode. The color code indicates the height of points on the surface relative to a reference line.    }
		\label{fig:afm_new&used}
	\end{figure}

A first objective of this study is to determine under what conditions spreading or pinned bubbles occur, and to find out in particular whether this is affected by electrochemical parameters, such as the electrolyte composition and the current density. In this work, \textit{new electrode} refers to one not previously used in experiments, whereas \textit{used electrode} denotes an electrode subjected to experimentation over several hours. The surface of new electrodes produced in the cleanroom is very smooth (see \hyperref[{fig:afm_new&used}]{Figure \ref{fig:afm_new&used}}a and section \textcolor{blue}{S3} in SI). During the operation, small cracks or cavities can occur on the surface, e.g. due to bubble detachment \cite{Kou_2021, Song_2020, Cherian_2014} and from the cleaning procedures carried out between experiments, leading to a rougher and chemically heterogeneous surface vs. the initial electrode state. 
Atomic force microscopy (AFM) measurements were conducted on both new and used electrodes. 3D reconstructions of the surface topographies are presented in \hyperref[{fig:afm_new&used}]{Figure \ref{fig:afm_new&used}}, and corresponding roughness parameters are listed in Table \ref{tab:afm_new_used}. To accurately characterize surface characteristics,  three distinct portions (\SI{20}{\micro\meter} x \SI{20}{\micro\meter}) on each surface were scanned and one of the scanned areas is shown. Notably, we did not see a meaningful difference in roughness levels between different scanned areas on each electrode. The values of the roughness parameters listed in Table \ref{tab:afm_new_used} were taken as the average values of these three areas. As evident in \hyperref[{fig:afm_new&used}]{Figure \ref{fig:afm_new&used}}a, the new electrode's surface is predominantly smooth, with occasional inhomogeneities, likely attributable to dust particles. The measured mean roughness (Sa) of the areas scanned on this electrode is 0.28 nm with a standard error of 0.01 nm. On the other hand, the surface of the used electrode appears to be rougher than that of the new electrode. The mean roughness (Sa) of the scanned areas is measured as 0.93 nm with 0.02 nm standard error. A detailed explanation of the roughness parameters listed in Table \ref{tab:afm_new_used} can be found in ref (\kern-0.4em\citenum{Gadelmawla2002}).

\begin{table}[ht]
\begin{tabular}{@{}lll@{}}
\toprule
                                     & \multicolumn{1}{c}{\begin{tabular}[c]{@{}c@{}}New\\ Electrode\end{tabular}} & \multicolumn{1}{c}{\begin{tabular}[c]{@{}c@{}}\hspace{2mm}Used\\ Electrode\end{tabular}} \\ \midrule
RMS Roughness (Sq)  [nm]    & \hspace{2mm}0.37  $\pm$ 0.02                                                                               & \hspace{2mm}1.42 $\pm$ 0.12       \vspace{5px}                                                                \\
Mean Roughness (Sa)  [nm]   & \hspace{2mm}0.28  $\pm$ 0.01                                                                               & \hspace{2mm}0.93 $\pm $ 0.02                                                                      \vspace{5px}  \\
Average Surface Height [nm] & \hspace{2mm}1.33 $\pm$ 0.01                                                                                 & \hspace{2mm}5.34 $\pm$ 0.59                                                                      \vspace{5px}  \\
Maximum Height  [nm]           & \hspace{2mm}7.07 $\pm$ 1.91                                                                                & 53.33 $\pm$ 7.08                                                                      \\ \bottomrule
\end{tabular}
 \caption{Surface topography of the new and used platinum electrodes.}
 \label{tab:afm_new_used}
\end{table}

\begin{figure}[ht]
		\centering
		\includegraphics[width=0.75\columnwidth]{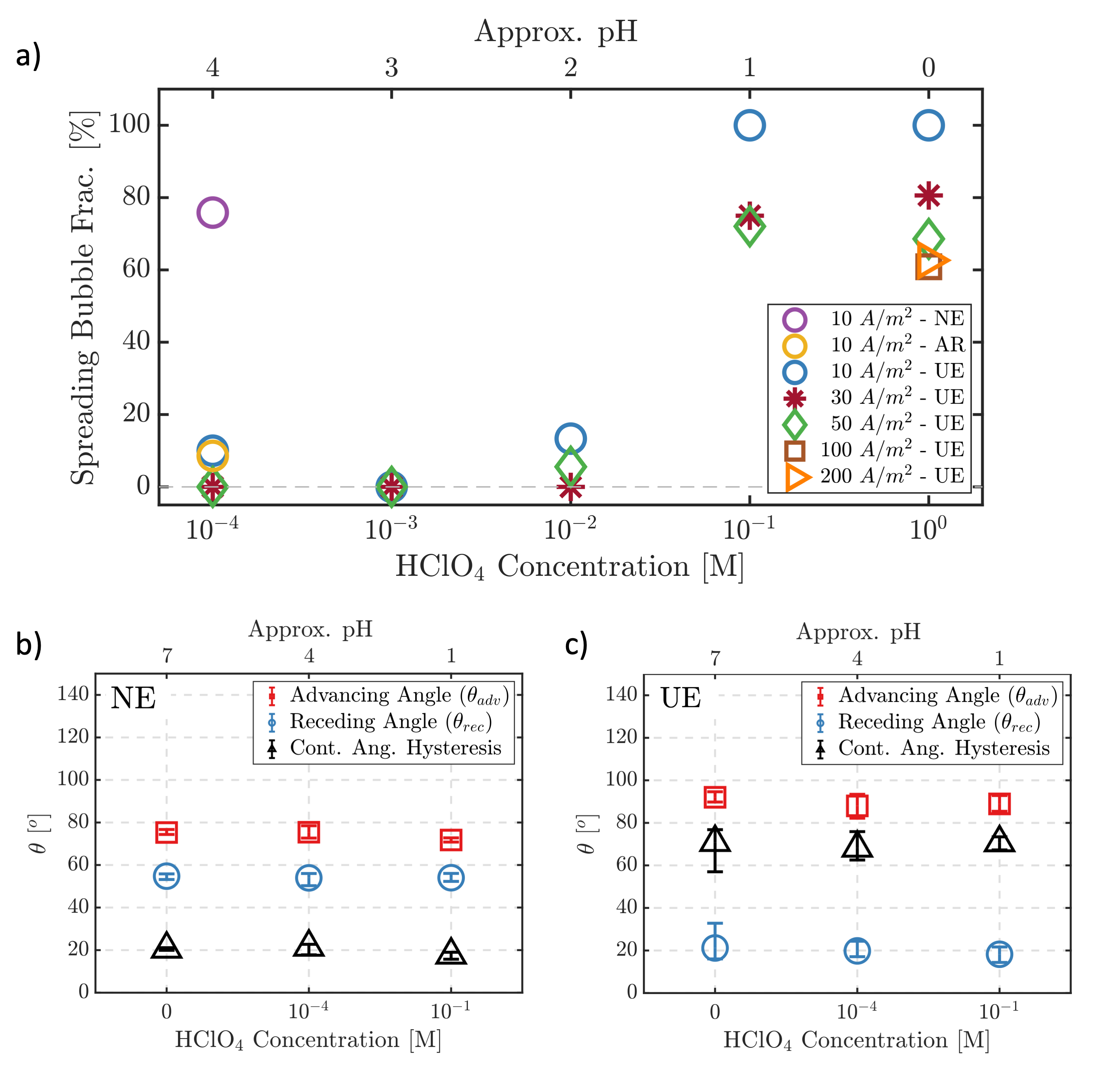}
		\caption{(a) The spreading bubble fraction as function of the $\ce{HClO4}$ concentration of the electrolytes at new electrode (NE), used electrode (UE), and after roughening the new electrode (AR). (b,c) Dynamic contact angles of drops from liquids with varying acidity measured by sessile drop experiments on a new and a used electrode, respectively. The data point at molar concentration of 0 was measured in deionised water. }
		\label{fig:spread_frac}
	\end{figure}

To check how the surface topography affects the bubble dynamics, we compared a new and a used electrode in $10^{-4}$ M of $\ce{HClO4}$ solution at a current density of 10 $\ce{A/m^2}$ (see \hyperref[{fig:spread_frac}]{Figure \ref{fig:spread_frac}a}). On the new electrode, most of the bubbles were observed to be spreading (purple circle). To accelerate the aging of the electrode, the current density was then increased to 200 $\ce{A/m^2}$ for 5 minutes to roughen the electrode. Repeating the experiment at 10 $\ce{A/m^2}$ after the roughening lead to a decrease in the fraction of spreading bubbles from $\sim$75\% on the new electrode to $\sim$9\% (yellow circle). More experiments with the same electrode did not further change the fraction of spreading bubbles significantly (blue circle). Therefore, considering that the initial smoothness of a new electrode is temporary, all data presented in the following are taken after preconditioning the electrodes at 200 $\ce{A/m^2}$ for 5 minutes or at a lower current density for a longer time.  
 
To determine the dependence of the spreading dynamics on the acid concentration, acid solutions in the range from $10^{-4}$ M to 1 M were investigated, evaluating at least 35 bubbles per solution. As \hyperref[{fig:spread_frac}]{Figure \ref{fig:spread_frac}}a shows, there is a strong dependence on the acid concentration, with pinned bubbles dominating at low concentrations $\leq 10^{-2}$ M, whereas spreading was found to be the prevalent mode at 0.1 M and beyond. The same figure also includes results for different current densities for each electrolyte concentration. It is seen that this parameter does not appear to affect the results substantially for the $\ce{HClO4}$ concentrations $\leq 10^{-2}$ M. At the larger concentrations, there is a slight trend towards lower spreading fractions with increasing current density, which is potentially related to the activation of additional nucleation sites.
	
To complete the picture, we performed contact angle measurements with different acid concentrations on a new and a used electrode (\hyperref[{fig:spread_frac}]{Figure \ref{fig:spread_frac}}(b,c)). The contact angle hysteresis is larger for the used electrode compared to the new one, indicating a significant change in surface properties. The corresponding results in \hyperref[{fig:spread_frac}]{Figure \ref{fig:spread_frac}}c show that, in particular, the advancing contact angle ($\theta_{adv}$) is mostly unchanged, while the receding contact angle ($\theta_{rec}$) has a small albeit noticeable decreasing tendency towards higher acid concentration. Based on these results, especially for $\theta_{rec}$, it appears unlikely that the change in the contact line dynamics observed in \hyperref[{fig:spread_frac}]{Figure \ref{fig:spread_frac}}a is related to variation of the contact angles. Instead, a possible explanation for the change in the contact line dynamics with the electrolyte concentration could be related to the surface charge of the bubbles. The isoelectric point corresponds to the pH value at which the zeta potential of a molecule or surface becomes zero, which occurs at around pH = 1.5-3 for gas bubbles, depending on the concentration and composition of the electrolyte \cite{Kosmulski2012, Nguyen_2003, Yang2001, Healy2007}. Therefore, bubbles exhibit negative charge at around pH \textgreater 2-3 and a positive charge at lower pH values than this range \cite{Brandon_1985, Kelsall_1996}. Given that the bubbles considered here form on the cathode of the cell, they are attracted to the electrode surface in the solutions with pH \textless 2, but repelled in cases with higher pH. This transition around pH = 2 is consistent with the pH-value at which we observed the change in spreading dynamics in  \hyperref[{fig:spread_frac}]{Figure \ref{fig:spread_frac}}a. It should be noted that the potentials applied in the experiments are in the range of $-0.36$ to $-3.2$ V (vs Ag/AgCl), and much lower than those typically considered for electrowetting, which are of the order of 10 - 120 V \cite{Mugele2005}. Therefore, we do not expect  electrowetting to have a significant influence here.

\subsection{Pinned Bubble Detachment}
\begin{figure}[ht]
		\centering
		\includegraphics[width=0.7\columnwidth]{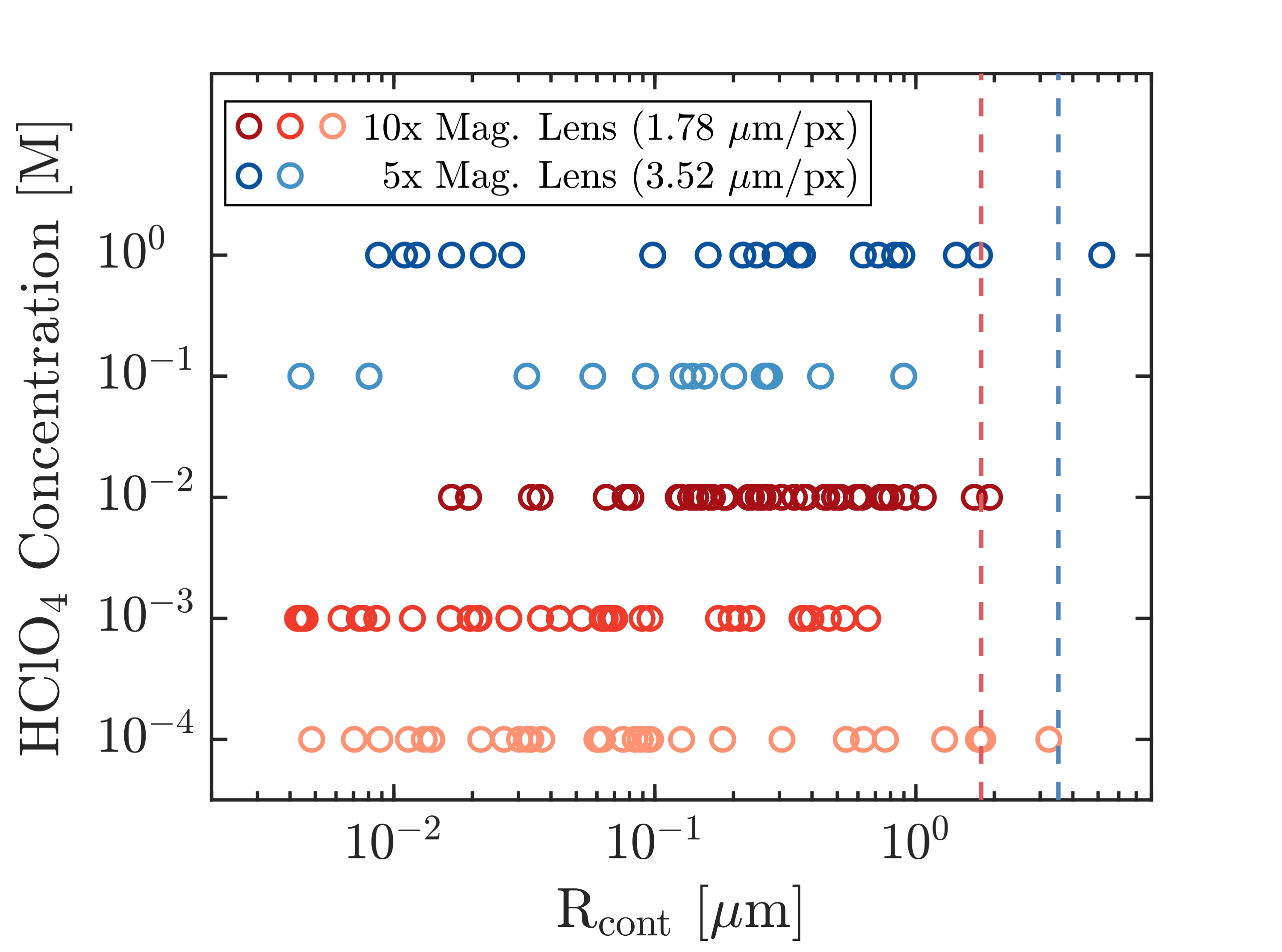}
		\caption{Contact radius ($R_{cont}$) of the pinned bubbles calculated by equation (\ref{eq:cav_det}). Red circles show the pinned bubbles detected in the experiments with acid concentrations ranging from $\mathrm{10^{-4}}$ to $\mathrm{10^{-2}}$ M, using the 10x magnification lens. Blue circles represent the 5x magnification lens used in the experiments with concentrations of 0.1 and 1 M. The red and blue dashed lines indicate the corresponding size of one pixel for a 10x magnification lens and a 5x magnification lens, respectively.}
		\label{fig:cav_det}
	\end{figure}

As shown in \hyperref[{fig:setup_exp_images}]{Figure \ref{fig:setup_exp_images}}b, it is not possible to obtain contact line information of pinned bubbles with our optical configuration. However, we can estimate $R_{cont}$ based on the optical measurement of the detachment radius of the pinned bubbles ($R_{cav}$) using equation (\ref{eq:cav_det}), to provide an estimate on the size range of the pinning sites. The results in  \hyperref[{fig:cav_det}]{Figure \ref{fig:cav_det}} show a large spread, consisting of more than three orders of magnitude, but no noticeable dependence on the electrolyte concentration. Even for the largest estimates, $R_{cont}$ is of comparable order to the image resolution, which is why the contact line cannot be resolved for the pinned bubbles.

\subsection{Evolution of a Spreading Bubble with Dynamic Wetting}

\begin{figure}[ht]
		\centering
		\includegraphics[width=1\columnwidth]{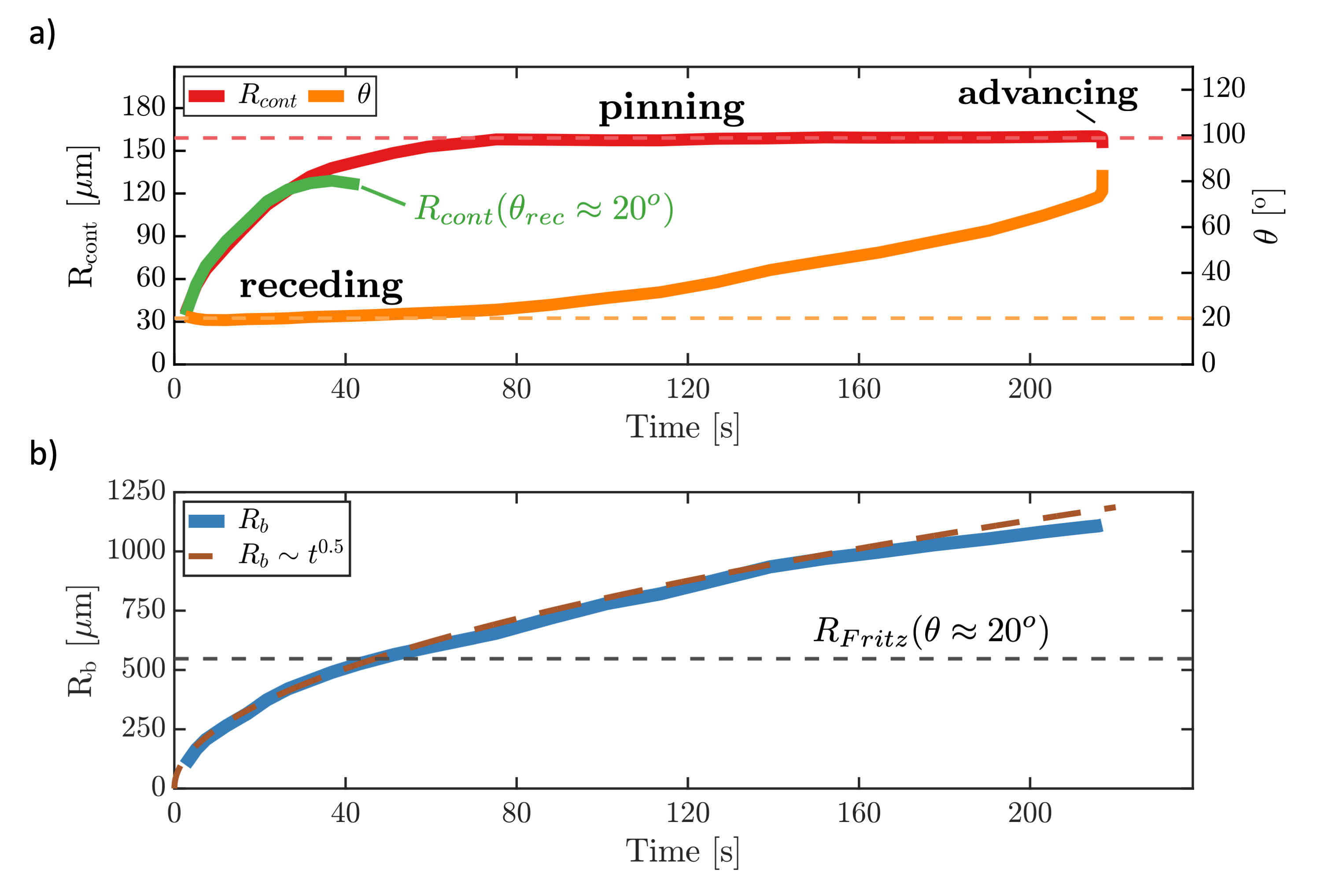}
		\caption{(a) Typical development of contact line radius ($R_{cont}$) and contact angle ($\theta$) during the growth of a spreading bubble. The green line indicates the expected evolution of the contact radius for constant $\theta$ consistent with the Fritz assumption. (b) The development of volume equivalent radius ($R_b$) of the same bubble during its growth. Dashed black line shows the $R_{Fritz}$ value calculated by equation (\ref{eq:Fritz}) at $\theta \approx 20 ^\circ$ }
		\label{fig:spread_evol}
	\end{figure}

A typical evolution of $R_{cont}$ (left ordinate) and $\theta$ (right ordinate) of a spreading bubble is depicted in \hyperref[{fig:spread_evol}]{Figure \ref{fig:spread_evol}}a. The corresponding evolution of $R_b$ is shown in \hyperref[{fig:spread_evol}]{Figure \ref{fig:spread_evol}}b. The bubble growth is typically characterized by a power law $\mathrm{R_b \sim t^\alpha}$, where $t$ represents the time from the onset of the nucleation, and the exponent $\mathrm{\alpha}$ depends on the relevant growth dynamics\cite{Brandon_1985, Yang_2015}. 
\hyperref[{fig:spread_evol}]{Figure \ref{fig:spread_evol}}b shows that $R_{b}$ of the bubble growing on a large electrode roughly follows a $\mathrm{\sim t^{0.5}}$ trend (brown dashed line), indicating diffusion-controlled growth ($\mathrm{\sim t^{0.5}}$) \cite{Glas1964, van_der_Linde_2017, Raman_2022}. As the bubble grows, the contact line initially spreads across the electrode surface while maintaining approximately constant $\theta\approx 20^\circ$ (receding phase). Consistent with the receding (dewetting) motion of the contact line during this period this angle is close to the values measured for $\theta_{rec}$ (see \hyperref[{fig:spread_frac}]{Figure \ref{fig:spread_frac}}b). 
After the initial spreading phase, $R_{cont}$ reaches a plateau (after about 80s in \hyperref[{fig:spread_evol}]{Figure \ref{fig:spread_evol}}a), i.e. the contact line gets pinned and does not spread anymore (pinning phase). At the same time, the measured $\theta$ increases up to a value of $\theta  \approx 70^\circ$, after which the bubble departs. Movie S1 shows these three phases of a bubble during the growth (see the Supporting Information). 
 
The green line in \hyperref[{fig:spread_evol}]{Figure \ref{fig:spread_evol}}a illustrates the development of $R_{cont}$ expected in the Fritz model, which assumes a constant contact angle throughout the entire bubble lifetime. Crucially, $R_{cont}$ for a bubble with constant $\theta$ reaches a maximum before the maximum bubble volume (at the end of the green line) is reached. This implies a change from receding (the contact patch spreading) to advancing (contact patch shrinking) contact line motion before departure of the bubble. Due to the contact angle hysteresis, the contact line in the experiment gets pinned at this transition and the actually measured (red) contact line starts to deviate from the one expected for a constant contact angle. A detailed investigation on the mechanisms of the pinning-depinning process of the contact line is beyond the scope of this work. Nonetheless, our finding appears consistent with studies reporting that even nanometer-scale structures can alter surface wettability and significantly affect gas-liquid contact line dynamics\cite{ramos_2003, ramos_2006, ramiasa_2013, franiatte_2022, delmas_2011}. 

\begin{figure}[ht]
		\centering
		\includegraphics[width=0.9\columnwidth]{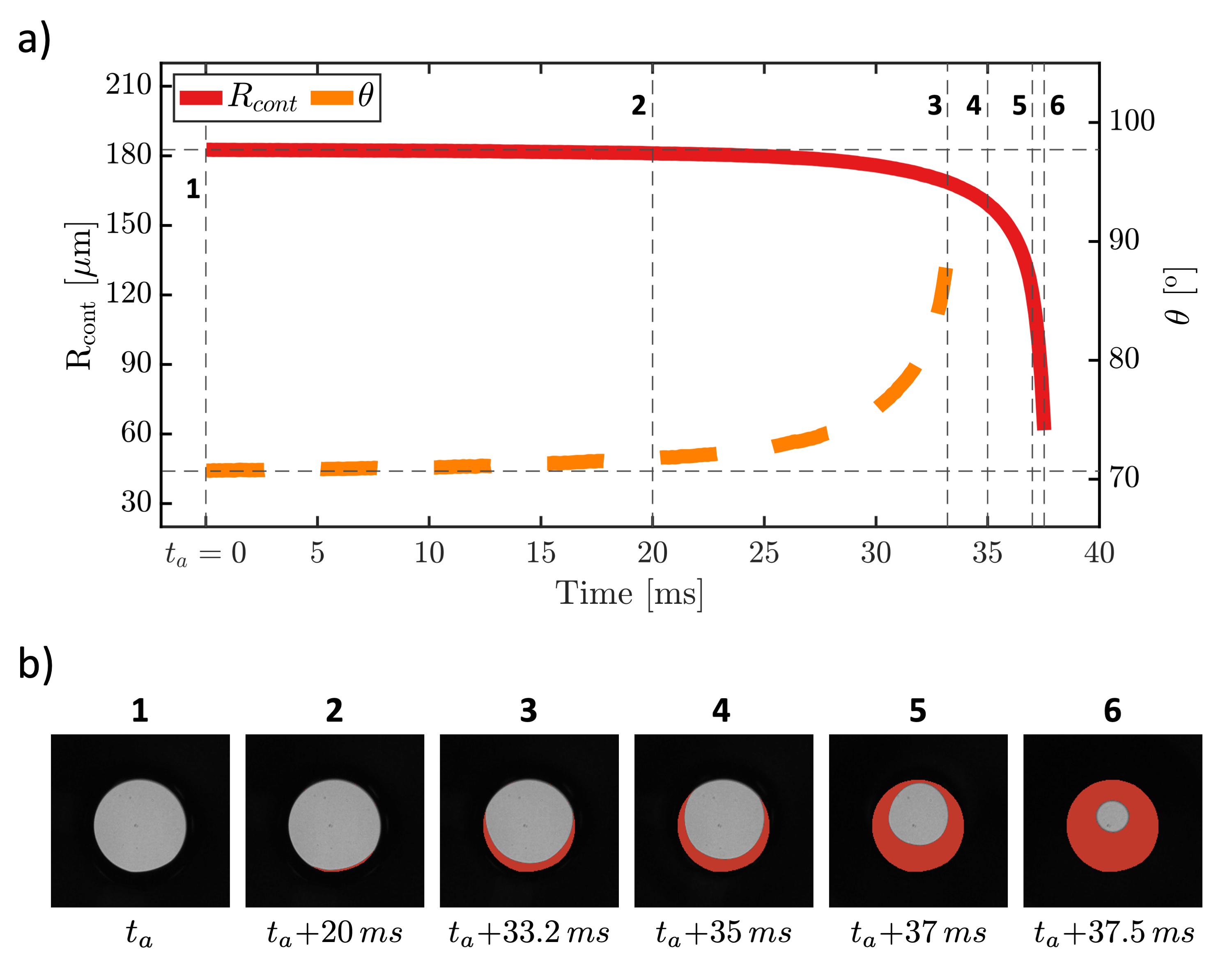}
		\caption{(a) Contact line radius $\mathrm{R_{cont}}$ and contact angle $\mathrm{\theta}$ evolution of a bubble in advancing stage. Left and right y-axes are for $\mathrm{R_{cont}}$ and $\mathrm{\theta}$, respectively. (b) Experimental snapshots corresponding to the vertical dashed lines in (a). Gray areas represent the contact line region, whereas the red areas indicate the change in the contact patch compared to time $t_a$. }
		\label{fig:adv_CL}
	\end{figure}

The bubble then continues to grow in the pinned stage. During this continued growth, the contact area now remains constant while the contact angle increases. This increase proceeds until the advancing contact angle is reached, at which point the contact line begins to move inwards. To capture the advancing motion of the contact line, which happens on a much shorter timescale compared to the bubble lifetime, we utilized high-speed imaging at 15,000 Hz. The evolution of $R_{cont}$ and $\theta$ of a bubble just before detachment, along with the corresponding snapshots of the contact line region (gray area) are shown in \hyperref[{fig:adv_CL}]{Figure \ref{fig:adv_CL}} (see also Movie S2 in Supporting Information). The contact line starts to advance at $\mathrm{t=t_a}$ and moves gradually until $\mathrm{t=t_a+20}$ ms. Subsequently, both the advancement of the contact line and the increase in $\theta$ accelerate significantly, resulting in the detachment of the bubble at $t \approx t_a + 38$ ms. 
Note that we have included ${\theta}$ as a dashed line here, since it is not clear if the force balance implied by the underlying solution of the YL-equation is still valid at this stage. That is, the bubble may already be accelerating upwards. This is definitely the case beyond $t \approx t_a + 33$ ms, where $\theta$ estimated based on the YL solution approaches $\approx \ $90 $^\circ$ (see \hyperref[{fig:adv_CL}]{Figure \ref{fig:adv_CL}a}), and no consistent solution of equation (\ref{eq:YL1}) exists at later times. Also the Reynolds number $Re = U_{cl}R_{cont}/{\nu}$, which is lower then 0.1 in the growth phase, exceeds 1 at later times $t \gtrapprox t_a + 33$, indicating that inertial effects become relevant at this stage. Here, $U_{cl} = dR_{cont}/dt$ is the contact line velocity and $\nu$ is the kinematic viscosity of the liquid. Similarly, the capillary number \(\mathrm{Ca = {\mu} U_{cl}/{\sigma}}\) (where $\mu$ is the dynamic viscosity of the liquid), for which typical values lie within the range $\mathcal{O}(10^{-8} - 10^{-9})$ during the spreading phase, increases by several orders of magnitude up to $\mathcal{O}(10^{-3})$ at the end of the advancing phase. Therefore, motion induced modifications to the contact angle potentially become relevant before detachment\cite{Snoeijer_2013}. Crucially though, with an overall duration of about 40 ms the advancing phase is very short compared to the overall growth time. Additional bubble growth during this period is therefore negligible and estimating the departure size at the start of the advancing phase is valid.

\begin{figure}[ht]
		\centering
		\includegraphics[width=0.7 \columnwidth]{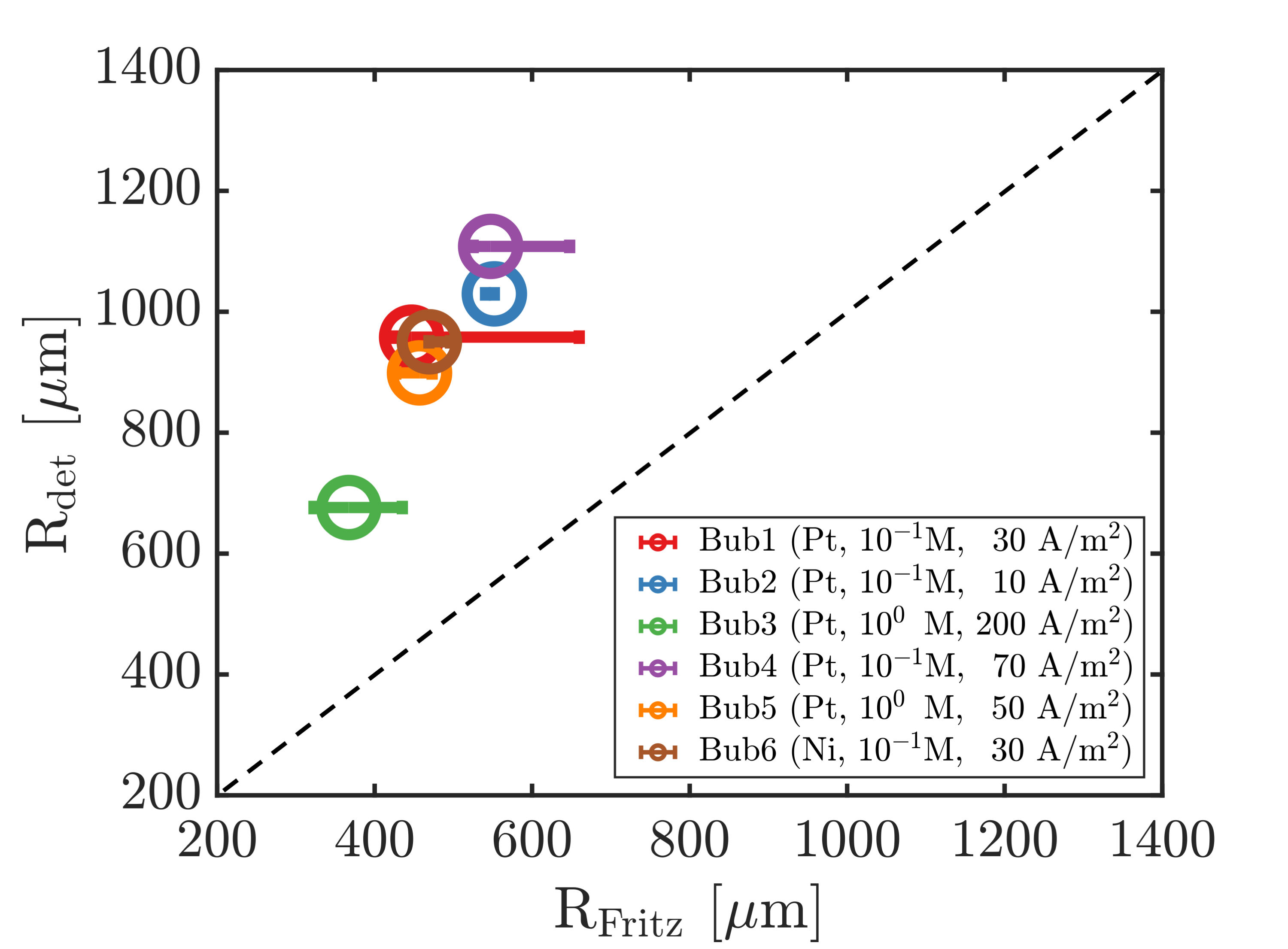}
		\caption{Experimental detachment radius $\mathrm{R_{det}}$ compared to the calculated Fritz radius $\mathrm{R_{Fritz}}$. The dashed line corresponds to the diagonal and serves as a reference.
    }
		\label{fig:R_FritzRdet}
	\end{figure}

As \hyperref[{fig:spread_evol}]{Figure \ref{fig:spread_evol}}a shows, the departure occurs at more than twice the departure radius (and therefore more than 8 times the volume!) compared to the Fritz prediction using the receding contact angle of $\theta_{rec} = 20^\circ$. Consistent results were observed for other bubbles in various experimental conditions as presented in \textcolor{blue}{Figure S4} of the supporting information. In \hyperref[{fig:R_FritzRdet}]{Figure \ref{fig:R_FritzRdet}}, we compare the measured departure radii with the respective Fritz prediction using the receding contact angle observed during the initial spreading phase of the bubble evolution. Horizontal error bars represent the uncertainty in the determination of $\theta_{rec}$ due to slight variations during the spreading. These results clearly show that the Fritz model is not well suited to predict the departure of bubbles on surfaces with contact angle hysteresis. 
The Fritz model describes bubble departure for an ideal case with $\theta_{rec} = \theta_{adv} = \theta_{eq}$, i.e. in the absence of contact angle hysteresis. However, our data  clearly shows that in the presence of contact angle hysteresis the receding phase is followed by a pinning phase. This is in contrast to the Fritz model, which predicts a short advancing stage followed by detachment. As a result of the pinning, the departure radius of the bubble and hence its `lifetime' are significantly extended compared to the Fritz prediction. These experimental findings are further in line with an earlier numerical study by \citet{Allred_2021} in the context of boiling, who also emphasised the need to consider dynamical wetting properties in predicting the bubble departure size on surfaces. 

\subsection{Force Balance and Detachment Radius with Dynamic Wetting}

\begin{figure}[ht]
		\centering
		\includegraphics[width=0.7\columnwidth]{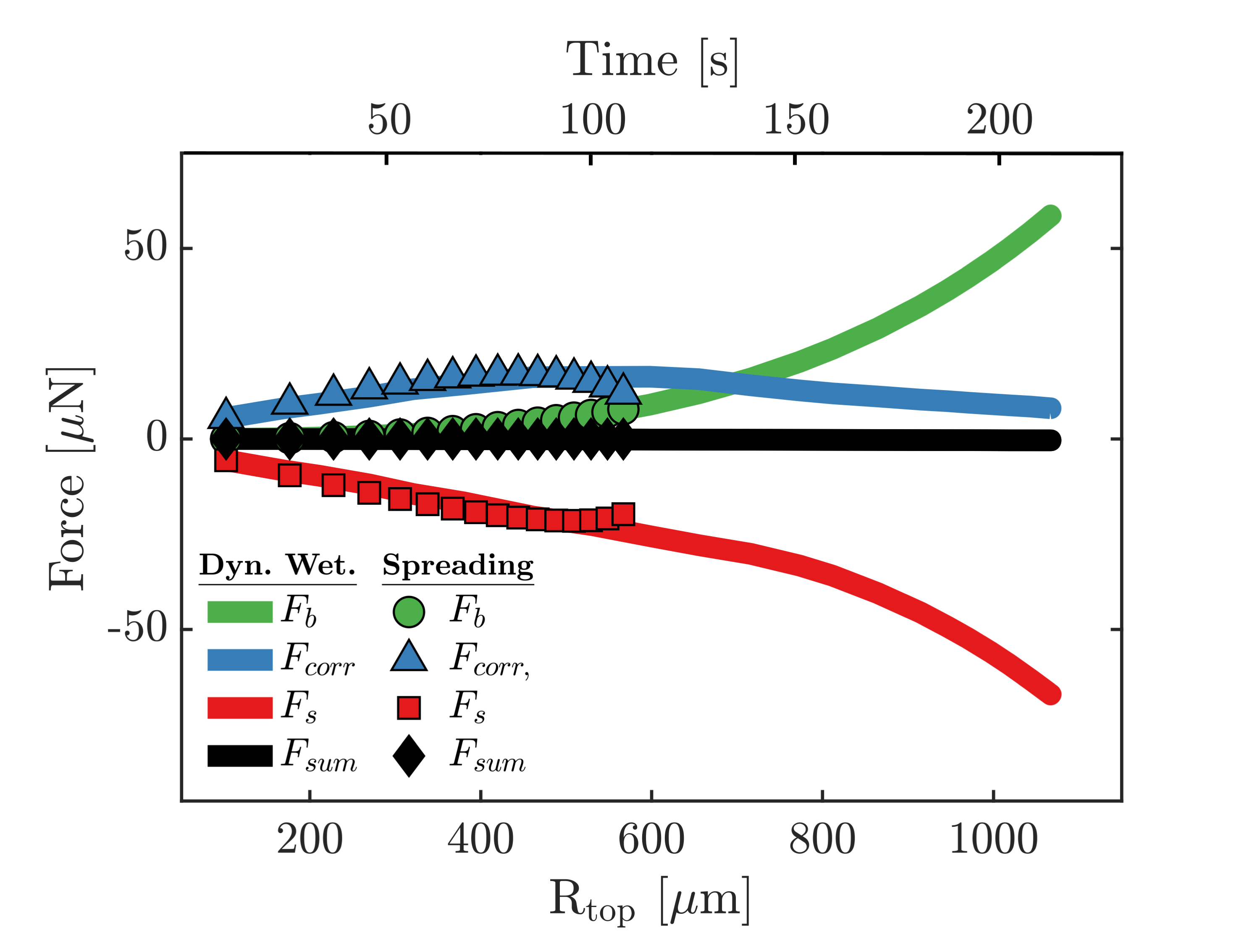}
		\caption{Force evolution of the experimentally observed spreading bubble (Bubble 4 in figures 8 and 10)}
		\label{fig:forces_real}
	\end{figure}

\hyperref[{fig:forces_real}]{Figure \ref{fig:forces_real}} shows the force balance for the same bubble as in \hyperref[{fig:spread_evol}]{Figure \ref{fig:spread_evol}}. As expected, in the early stages of development the force balance equals that of a spreading bubble (shown as markers) with the dominant balance between $F_{corr}$ and $F_s$. After the pinning on the contact line at an approximate bubble size of $R_{top} = \SI{600}{\micro\meter}$, the pressure force $F_{corr}$ decreases in magnitude somewhat more slowly than for the $\theta = \textrm{const.}$ case since the radius of the contact patch does not decrease here. Nevertheless, buoyancy ($F_b$) becomes the dominant detaching force already at $R_{top}  \approx \SI{700}{\micro\meter}$ and exceeds $F_{corr}$ significantly beyond that. At detachment, $F_b \approx 10 F_{corr}$, such that the effective balance $F_b \approx F_s$ is equal to that of a pinned bubble, although with $R_{cont} \approx \SI{160}{\micro\meter}$ the contact patch in the present case is much larger than for typical pinned bubbles with $R_{cont} \leq \SI{10}{\micro\meter}$ (see \hyperref[{fig:cav_det}]{Figure \ref{fig:cav_det}}). Since detachment occurs when the contact angle of the pinned contact line reaches $\theta_{adv}$, we can state the force balance at detachment as
\begin{equation}
    4/3\pi R_{det}^3\rho_l g \approx  2\pi R_{cont} \sigma \sin \theta^*_{adv}.
    \label{eq:FBd}
\end{equation}
This approximation was also used in ref (\kern-0.4em\citenum{Allred_2021}); note that $\theta^*_{adv} = \theta_{adv}$ for $\theta_{adv}<90^\circ$, but since attachment will occur at this angle, the value is limited to $\theta^*_{adv}= 90^\circ$ if $\theta_{adv}\geq 90^\circ$. 

Based on equation (\ref{eq:FBd}), it is possible to determine the detachment radius $R_{det}$ of the bubble, provided the size of the contact patch is known. The relevant value $R_{cont,max}$ is determined by the maximum patch radius during the initial spreading with $\theta_{rec}$. For the case of small Bond numbers ($\mathrm{Bo \leq 0.1}$), which is typically applicable here, Chesters \cite{Chesters_1978} derived an analytical solution which gives the maximum patch radius as

\begin{equation} 
\frac{R_{cont,max}}{\lambda_c} = \frac{9}{32}{\sqrt{2}}\sin^2\theta_{rec}.
\label{eq:maxR_cont}
		\end{equation}

\begin{figure}[ht]
		\centering
		\includegraphics[width=1\columnwidth]{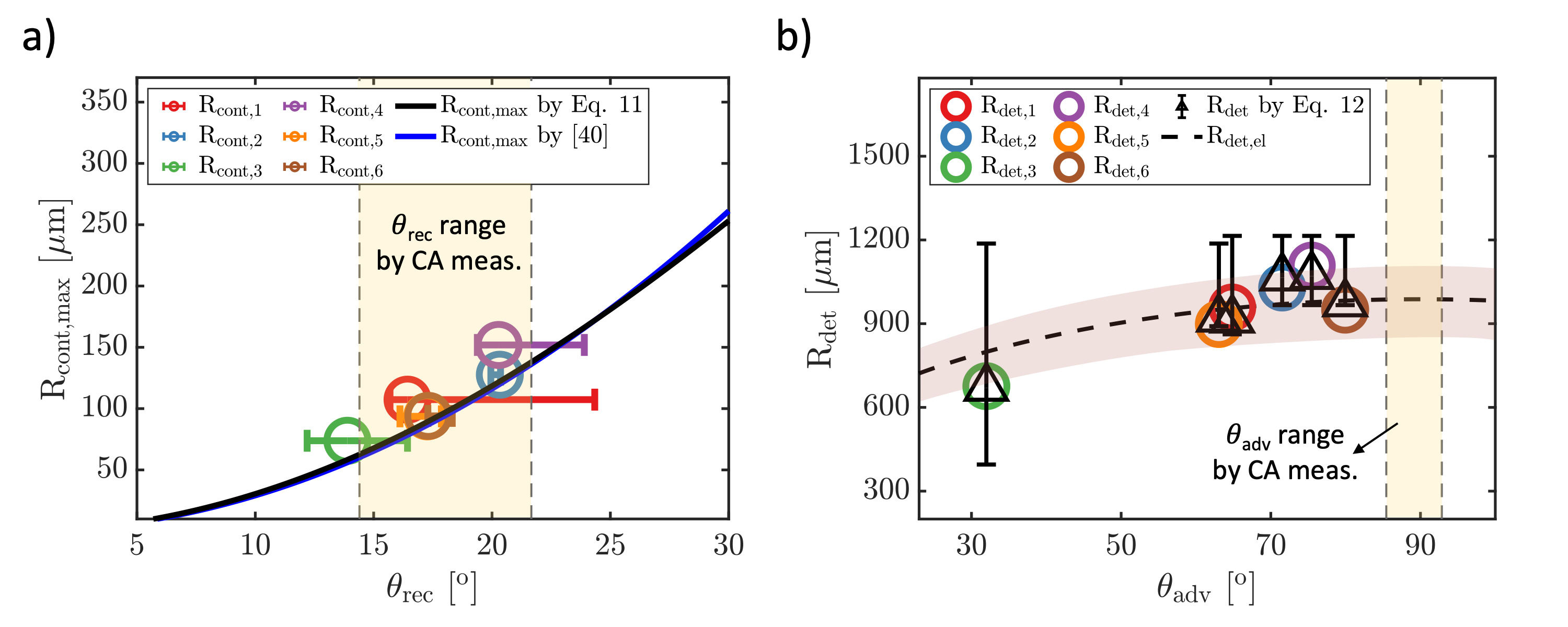}
		\caption{The estimation of (a) the maximum contact line radius ($R_{cont, max}$), and (b) detachment radius ($R_{det}$) by receding ($\theta_{rec}$) and advancing ($\theta_{adv}$) contact angles. The red colored area in (b) indicates the characteristic $R_{det}$ range of a used electrode. $\theta_{rec}$ and $\theta_{adv}$ were measured by sessile drop experiments, and $R_{det}$ was calculated using equation (\ref{eq:R_det}).}
		\label{fig:R_Cont - Rdet}
	\end{figure}

\noindent Considering that there is some uncertainty in the determination of $\theta_{rec}$, this relationship between $R_{cont}$ and $\theta_{rec}$ is found to be consistent with our data as shown in  \hyperref[{fig:R_Cont - Rdet}]{Figure \ref{fig:R_Cont - Rdet}}a. The figure also includes an empirical relationship for $R_{cont,max}$ provided in ref (\kern-0.4em\citenum{Allred_2021}) based on their simulation results, which closely agrees with equation (\ref{eq:maxR_cont}).

From equations (\ref{eq:FBd}) and (\ref{eq:maxR_cont}) it becomes clear that the receding contact angle determines the size of the contact patch while the end of the pinning phase and hence detachment depends on $\theta^*_{adv}$. By combining the two equations, we arrive at the following relationship between $\theta_{rec}$ and $\theta_{adv}$ and $R_{det}$:

\begin{equation} \label{eq:R_det}
R_{det} = \frac{3}{4}{{\lambda}_c}\biggl(\sqrt{2}\sin^2{\theta_{rec}}\sin{\theta^*_{adv}}\biggl)^\frac{1}{3}.
		\end{equation}

\begin{figure}[ht]
		\centering
		\includegraphics[width=1\columnwidth]{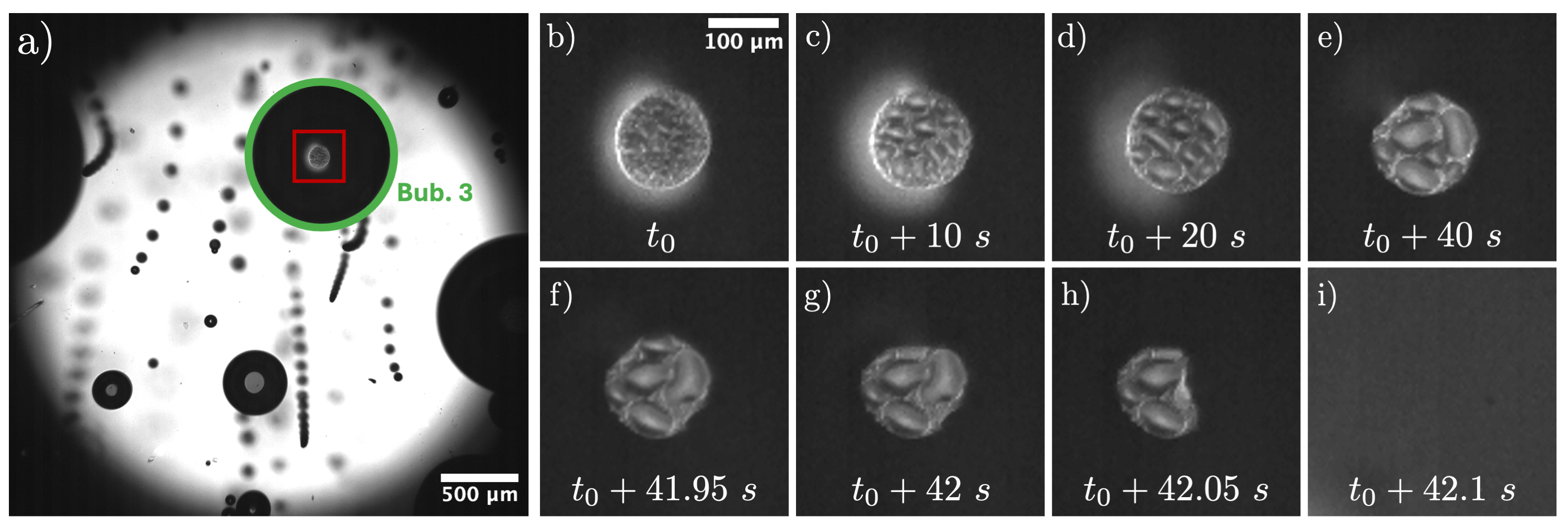}
		\caption{(a) Snapshot of the bubble population during evolution of bubble 3 (green circle in figures \ref{fig:R_FritzRdet} and \ref{fig:R_Cont - Rdet}). (b-i) Time sequence of the contact patch (red rectangle in panel (a)) during the evolution of the bubble (b-e) and shortly before departure (f-i).
  }
		\label{fig:Drop}
	\end{figure}

\noindent In \hyperref[{fig:R_Cont - Rdet}]{Figure \ref{fig:R_Cont - Rdet}}b, equation (\ref{eq:R_det}) is compared to the  experimentally obtained values. We make this comparison on two levels. First, we measure the receding contact angle from the bubble evolution (as in \hyperref[{fig:spread_evol}]{Figure \ref{fig:spread_evol}}) and use the contact angle at the moment when the contact line starts advancing as $\theta^*_{adv}$. The corresponding results are shown as black triangle markers and conceptually validate equation (\ref{eq:R_det}).  A more practical comparison is to use the contact angles obtained from the sessile drop experiments (See section S4 in the supporting information for the methodology of the technique). Again, these results (shown as red shading) are in good agreement with the measured detachment radii. The only exception is bubble 3, for which the contact angle at detachment is significantly lower compared to the other bubbles, resulting in a lower value $R_{det}$. To understand this, a snapshot of the bubble population is shown in \hyperref[{fig:Drop}]{Figure \ref{fig:Drop}} together with a time series of the contact area evolution for this bubble.

Bubble 3 is observed in the experiment with the highest nominal current density of $\mathrm{200~A/m^2}$. This leads to a very dynamic bubble population with many microbubbles detaching as shown in the snapshot in \hyperref[{fig:Drop}]{Figure \ref{fig:Drop}}a. It is known\cite{Fernandez_2014} and also shown conclusively for the present case \cite{Bashkatov2024, Sanjay2024}, that coalescence with such microbubbles can lead to droplet injection into this bubble. \hyperref[{fig:Drop}]{Figures \ref{fig:Drop}}(b-e) show that such droplets accumulate on the electrode within the contact patch over time, filling increasingly more space and forming larger droplets. If such a large droplet merges with the contact line of the bubble, as is the case for bubble 3 around $t_0+42s$ (see \hyperref[{fig:Drop}]{Figures \ref{fig:Drop}}(f-i)), this can lead to a sudden and substantial reduction of the contact area. In the case of bubble 3, this is sufficient to induce bubble departure before reaching $\theta = \theta_{adv}$, as would be expected. It should be noted that also the other bubbles depart slightly before reaching the range of $\theta_{adv}$ determined from the sessile drop experiments. It remains unclear, whether this is for the same reason or if other factors, such as the fact that the force balance underlying equation (\ref{eq:FBd}) is only approximate, play a role. 

\section{CONCLUSIONS}
We have experimentally investigated the connection between contact line dynamics and buoyancy-driven bubble departure during water electrolysis. We observed a significantly reduced probability of contact line spreading in electrolytes with lower acid concentrations, while contact line spreading was more likely for acid concentrations of $10^{-1}$ M and higher. The absence of a noticeable variation in the contact angle in this pH range suggests a change in the surface charge of the bubbles as a potential cause of this effect. Observed departure sizes of pinned bubbles, i.e. without contact line spreading, imply typical contact patch radii $R_{cont} \leq \SI{1}{\micro\meter}$, which cannot be resolved in the present experimental configuration. For spreading bubbles, we find that our experimental results for the departure radius do not agree with the widely used 'Fritz radius' \cite{Fritz_1935}. Our data reveal that the reason why the bubbles in the experiment remain attached to the electrode for much longer than predicted by ref (\kern-0.4em\citenum{Fritz_1935}) is related to contact line hysteresis. This leads to pinning of the contact line after the initial spreading of the patch with $\theta_{rec}$ until the advancing contact angle $\theta^*_{adv}$ is reached, followed by the departure of the bubble. Similarly to what was found in ref (\kern-0.4em\citenum{Allred_2021}), we find that the pinned contact radius for these bubbles is equal to the maximum patch size possible for spreading with $\theta = \theta_{rec}$. The departure at the end of the pinning phase determined by $\theta \approx \theta^*_{adv}$ is characterised by an approximate equilibrium between surface tension and buoyancy. A prediction of the departure radius based on these results is found to be in good agreement with the experimental data. This agreement is also an indication that in our system other contributions by e.g. electric or Marangoni forces only play a secondary role in determining the detachment diameter. Interestingly, we also observe that pre-wetting of the contact patch, presumably due to droplets generated during coalescence with smaller bubbles, can lead to earlier departure of spreading bubbles. This effect is expected to be more prevalent at higher current densities where bubble coalescence is more frequent.

\section{Supporting Information}

Derivation of the force balance and sketch of the forces acting on the bubble (Figure S1); Electrolyte properties (Table S1); Surface characterization details and scanning electron microscopy (SEM) images (Figure S2); Details of the wetting characteristics and equilibrium contact angle of a new and used electrode (Figure S3); Contact radius and contact angle evolution of the spreading bubbles at various experimental conditions (Figure S4). 

\begin{acknowledgement}

This work was supported by the Dutch Research Council (NWO) in the framework of the ENW PPP Fund for the top sectors with support from Shell, Nobian and Nouryon [and from the Ministry of Economic Affairs in the framework of the PPS-toeslagregeling], Grant No. 741.019.201. DK has received funding from the European Research Council (ERC) under the European Union’s Horizon 2020 research and innovation programme (grant agreement No. 950111, BU-PACT). We thank Andrea Prosperetti, Bastian Mei and Aleksandr Bashkatov for fruitful discussions and their comments on the work, Kai Sotthewes for the AFM measurements, and the technicians of our research group for their great help on the design and production stage. We also thank the anonymous reviewers for their insightful questions and suggestions to improve our work. 
 
\end{acknowledgement}

\bibliography{main}

\end{document}


\tableofcontents
\newpage
\setcounter{figure}{0}
\setcounter{equation}{0}

\makeatletter 
\renewcommand{\thefigure}{S\@arabic\c@figure}
\renewcommand{\thetable}{S\@arabic\c@table}
\renewcommand{\thesection}{S\@arabic\c@section}
\renewcommand{\theequation}{S\@arabic\c@equation}

\makeatother


\section{Force Balance on a Bubble}
\label{SI_force_balance}

In deriving the force balance, we restrict ourselves to the effects of surface tension, pressure and buoyancy. Other factors, such as thermal\cite{Meulenbroek_2021, Massing_2019} or solutocapillary \cite{Park_2023, Park2024} Marangoni are also known to play a role, but are likely to be less relevant in the present case due to the low current densities in the experiment. The bubble growth is slow and occurs in the absence of external flow, such that inertial or other hydrodynamic forces are negligible. Note that we also neglect a possible contribution of an electric force, mainly because the wide range of reported surface charge density estimates in the literature \cite{Meulenbroek_2021, Bashkatov_2019} does not allow for a definitive evaluation of this contribution. 
\begin{figure}[h]
		\centering
		\includegraphics[width=1\columnwidth]{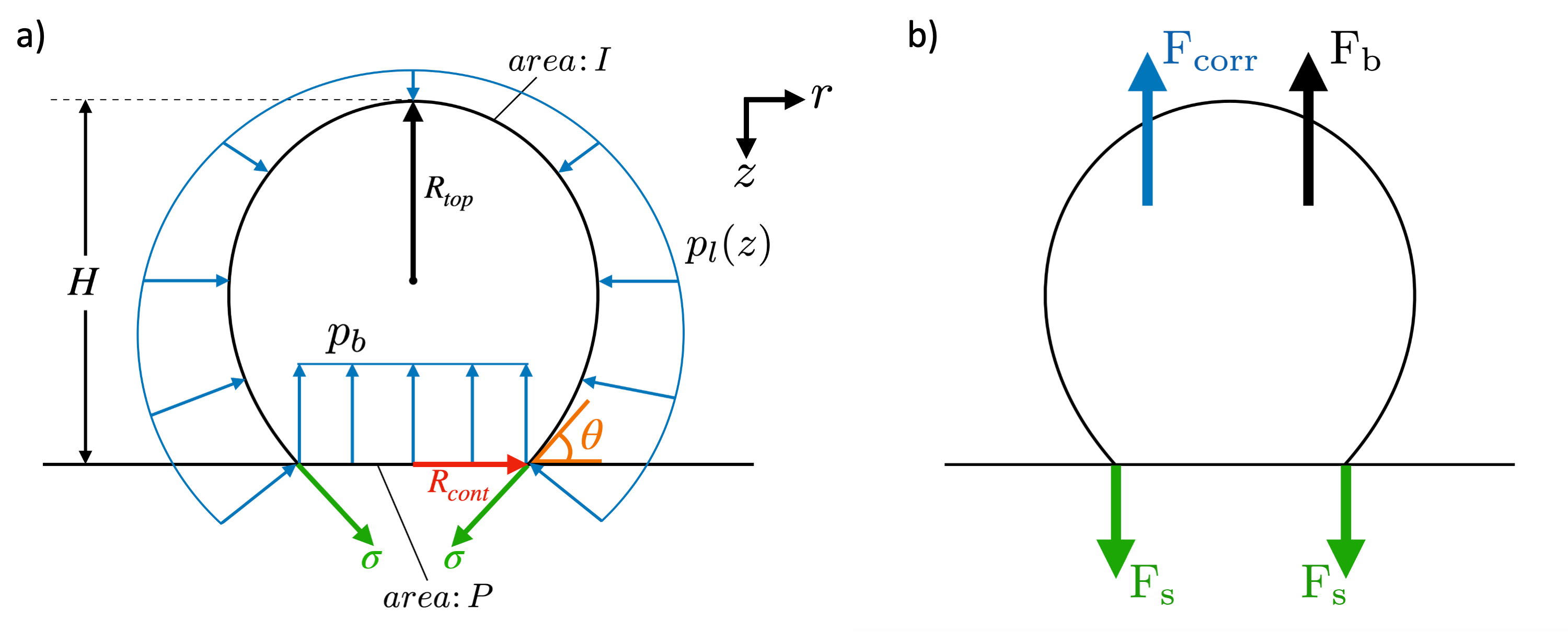}
		\caption{Schematics of the (a) stresses and (b) main forces acting on a bubble during growth. }
		\label{sfig:force_sketch}
\end{figure}

We consider a bubble in a fluid at rest for which the hydrostatic pressure distribution is given by 
\begin{equation}
    p_l(z) = p_l(z=0) + \rho_l g z,
\end{equation}
where $\rho_l$, g, and $z$ denote liquid density, gravitational acceleration, and the vertical distance from the origin line at the bubble top ($z = 0$), respectively (See Figure \ref{sfig:force_sketch}a). Applying Laplace's law at the top of the bubble, the pressure $p_b$ inside the bubble can be obtained from



\begin{equation}
    p_b - p_l(z=0) = \frac{2\sigma}{R_{top}},
    \label{eq:Laplace}
\end{equation}

with $\sigma$ denoting the surface tension and $R_{top}$ the radius of curvature at the top of the bubble. Note that since $\rho_g\ll \rho_l$, $p_b$ can be assumed to be constant within the bubble.

The resulting force in the vertical ($\vec{e_z}$) direction $F_z$ on the bubble follows from the integration
\begin{equation}
    F_z = \iint_I p_l \vec{e_z}\cdot \vec{n} dA+ \iint_P p_b \vec{e_z}\cdot \vec{n} dA - \oint_C \sigma \sin \theta dl.
\end{equation}
Here, "$I$" and "$P$" denote the surfaces (with normal vector $\vec{n}$) of the gas-liquid interface and the contact patch, respectively, and "$C$" is the contact line. Noting that the buoyancy force $F_b = \iint_{I+P} p_l(z) \vec{e_z} \cdot \vec{n} dA$ and using the surface tension force $F_s = \oint_C \sigma \sin \theta dl = 2\pi R_{cont} \sigma \sin \theta $, we obtain
\begin{equation}
    F_z = F_b + \iint_P (p_b -p_l(z))\vec{e_z}\cdot \vec{n} dA -F_s
\end{equation}
and
\begin{equation}
    F_z = F_b + {\pi}R_{cont}^2(p_b -p_l(H)) -F_s.
\end{equation}
Using equation (\ref{eq:Laplace}), the second term in the previous equation known as the `pressure correction force' $F_{corr}$ can be written as
\begin{equation}
    F_{corr} = \pi R_{cont}^2\left(\frac{2\sigma}{R_{top}}-\rho_l gH\right).
    \label{eq:correction}
\end{equation}
The correction force results from the fact that the expression for $F_b$ assumes hydrostatic pressure  at the bubble foot area $P$, whereas the actual pressure acting there is $p_b$. The first term in equation (\ref{eq:correction}) corresponds to the contact pressure force on the bubble foot while the second one represents a buoyancy correction. For a bubble to remain stationary, the resulting vertical force must be zero \emph{at all times}, which leads to the condition
\begin{equation}
    \frac{dv}{dt}= F_z   =0 = F_b + F_{corr} - F_s 
    \label{eq:static}
\end{equation}

\section{Electrolyte Properties}
\label{SI_electrolyte_propertie}

The electrolyte solutions used in the electrolysis experiments were prepared in different acid concentrations (from $10^{-4}$ M to 1 M). Furthermore, 0.5 M $\ce{NaClO4.H2O}$ was added into the solutions as supporting electrolyte (except 1 M $\ce{HClO4}$ case). In this way, the electrical conductivity of the solution was increased and high ohmic overpotentials were avoided during electrolysis, so that experiments up to a current density of 50 $\ce{A/m^2}$ could be performed for low acid concentrations. In this study, the bubble profile and contact angle were determined using the Young-Laplace (YL) equations, necessitating the calculation of the Bond number Bo = $\mathrm{{\Delta}{\rho}{g}{R_{top}^2}/{\sigma}}$. Therefore, physical properties of the electrolytes such as density and surface tension should be characterized. 

The density of the mixture solution was found by dividing the total mass of the solution by the total volume. The total mass and volume were calculated by summing the mass and volume of each substance, namely water, $\ce{HClO4}$ and $\ce{NaClO4.H2O}$. On the other hand, the surface tension of the electrolytes was found by pendant drop measurements. The experiments were carried out with an optical contact angle goniometer (OCA 15 Pro from Dataphysics Instruments), and instrument’s software (SCA20) was used to determine the drop shape and calculate the surface tension. Using a Hamilton syringe, a drop was generated on the tip of a needle (outer diameter of 0.718 mm) and pumped in very slowly (\SI{0.05}{\micro\liter}/s) until  it starts swing up and down. Subsequently, the corresponding surface tension value is found from the shape of the drop at that moment. Here, selecting a sufficiently large needle is crucial to enhance the accuracy of the measurements, as it ensures the formation of a large drop with significant deformation. The measurements were repeated with five drops for each liquids, and the average value were taken as ${\mathrm{\sigma}}$ value. The values of $\rho$ and $\sigma$ are shown in Table \ref{tab:rho_sigma}.

\begin{table}[ht]
\begin{tabular}{lll}
\toprule
& \shortstack[c]{$\rho$ \\ {[kg/m\textsuperscript{3}]}} & \shortstack[c]{$\sigma$\hspace{-1em} \\ \hspace{1em}{[mN/m]}} \\ 
\specialrule{\heavyrulewidth}{0pt}{1pt} 
Water & \hspace{2mm}1000 & 72.70 $\pm$ 0.09 \\
$10^{-4}$ M HClO\textsubscript{4} + 0.5 M NaClO\textsubscript{4} & \hspace{2mm}1035 & 70.23 $\pm$ 0.09 \\
$10^{-3}$ M HClO\textsubscript{4} + 0.5 M NaClO\textsubscript{4} & \hspace{2mm}1036 & 70.25 $\pm$ 0.09 \\
$10^{-2}$ M HClO\textsubscript{4} + 0.5 M NaClO\textsubscript{4} & \hspace{2mm}1036 & 70.19 $\pm$ 0.08 \\
$10^{-1}$ M HClO\textsubscript{4} + 0.5 M NaClO\textsubscript{4} & \hspace{2mm}1041 & 69.31 $\pm$ 0.07 \\
1 M HClO\textsubscript{4} &  \hspace{2mm}1057 & 67.22 $\pm$ 0.10 \\
\bottomrule
\end{tabular}
\caption{Density ($\rho$) and surface tension ($\sigma$) of the electrolytes used in the experiments at 20$^{\circ}$C }
\label{tab:rho_sigma}
\end{table}

\section{Surface Characterization}
\label{SI_surf_char}

Atomic force microscope (AFM) and scanning electron microscope (SEM) were employed to characterize the surface conditions of the new and used electrodes. Nanoscale topography images are acquired in tapping mode using an AFM (Bruker Icon) under ambient conditions with a humidity of approximately 48\% (measured with TFA Digital Professional Thermo-Hygrometer KLIMA BEE). A heavily doped n-type Si cantilever with a resonance frequency of 85 kHz and a force constant of 2.7 N/m (SSS-FMR, Nanosensors) was used. An open-source software (Gwyddion) was utilized for post-processing the raw images and extracting statistical values.

\begin{sidewaysfigure}[ht]
    \centering
    \includegraphics[width=0.9\columnwidth]{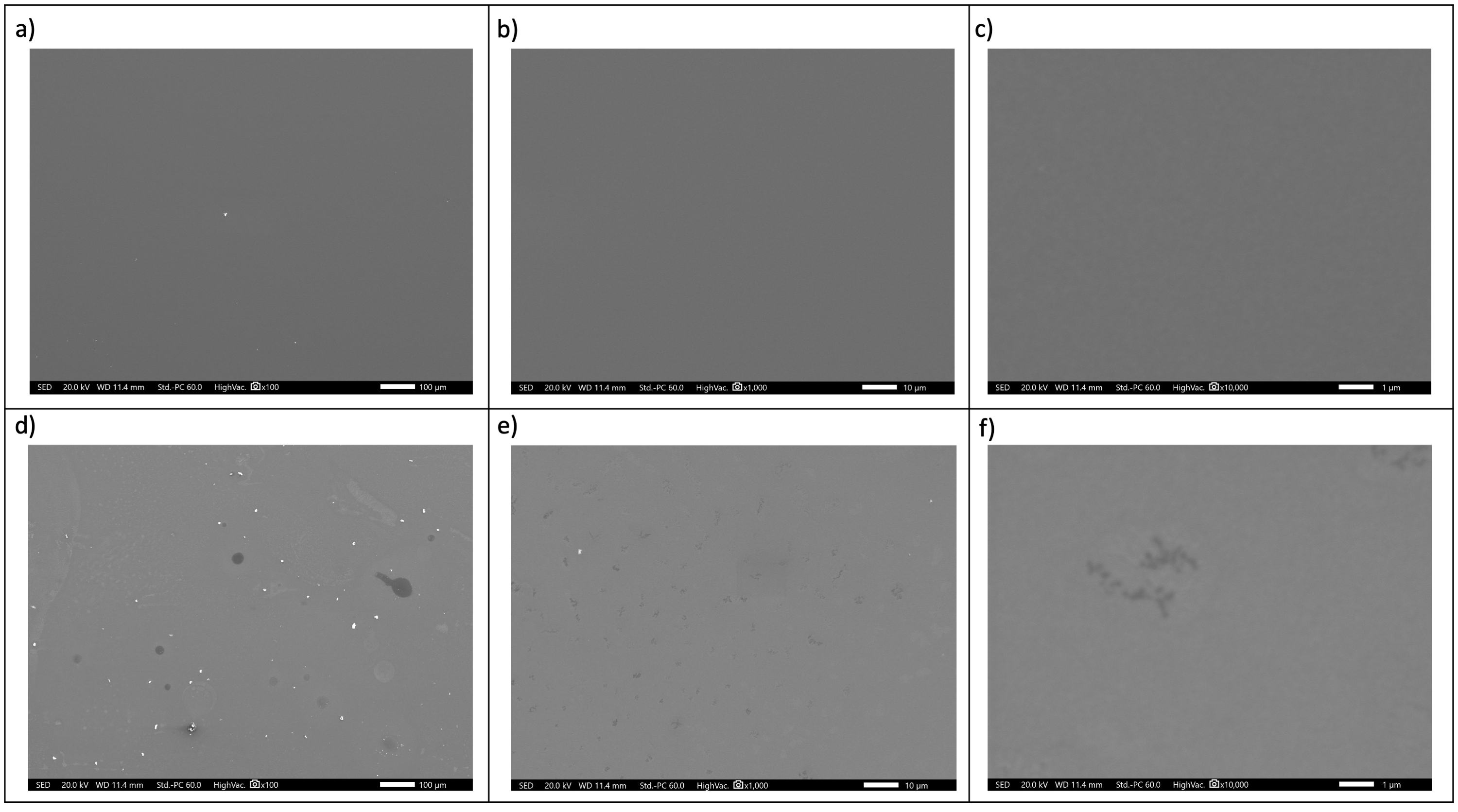}
    \begin{minipage}{0.9\columnwidth}
        \caption{Scanning electron microscope (SEM) images of a new and used electrodes. Images (a), (b), and (c) represent the surface of the new electrode in three magnification levels (with scale bars of \SI{100}{\micro\meter}, \SI{10}{\micro\meter}, and \SI{1}{\micro\meter}, respectively). On the other hand, images (d), (e), and (f) show the surface of a used electrode at the same scales.}
        \label{sfig:SEM}
    \end{minipage}
\end{sidewaysfigure}

\setlength\parindent{20pt} Additionally, the surface morphology of an electrode was observed using a SEM (JSM-IT200 from Jeol Ltd.) under the high vacuum environment, with magnifications up to 10,000x. The aperture was positioned at a distance of 11.4 mm from the substrate. The beam voltage and probe current (PC) were set to 20 kV and 60 nA, respectively. The SEM images of the new and used electrodes are shown in Figure \ref{sfig:SEM}, with magnification increasing from left to right within each row. Among the scanned areas on the new electrode, no remarkable surface features were seen, neither for a large area (a), nor a smaller area (c). Conversely, the damage on the electrode surface due the bubble detachment and surface cleaning between the experiments is evident at all magnification levels (d, e, f).

\newpage
\clearpage

 \newpage
 \clearpage

\section{Wetting Characteristics}
\label{SI_wet_char}

\begin{figure}[ht]
		\centering
		\includegraphics[width=0.6\columnwidth]{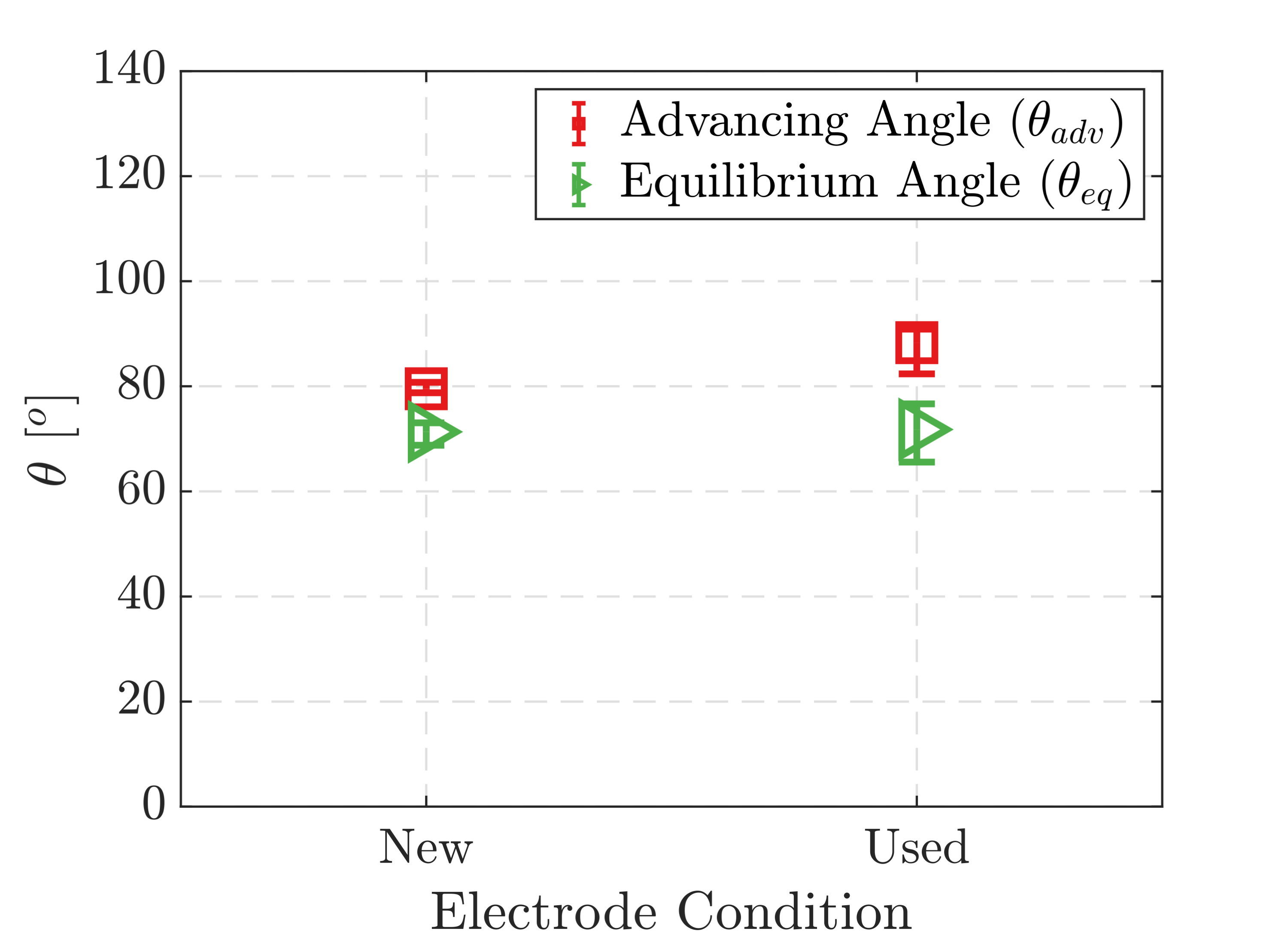}
		\caption{Advancing ($\theta_{adv}$) and equilibrium ($\theta_{eq}$) angles of water on a new and a used electrode, as determined through sessile drop experiments.}
		\label{sfig:equib_angle}
	\end{figure}

The wetting properties of the new and used electrodes were determined by sessile drop experiments. Similar to the pendant drop experiments, sessile drop experiments were also performed using an optical contact angle goniometer (OCA 15 Pro from Dataphysics Instruments), and a Hamilton syringe-needle pair. To determine the dynamic contact angles ($\theta_{adv}$ and $\theta_{rec}$) and equilibrium contact angle ($\theta_{eq}$) of the electrodes, two different techniques were employed. First, the dynamic contact angles of a new and a used electrode were measured using needle-in-drop sessile drop technique (see \textcolor{blue}{Figure 5}(b,c) in the main text). In this technique, $\theta_{adv}$ is measured as the contact line spreads across the surface during the deposition of a drop. Right after that, the deposited amount is withdrawn by the syringe, and $\theta_{rec}$ is measured as the contact line shrinks. In the second technique, $\theta_{adv}$ and $\theta_{eq}$ of a drop were measured. $\theta_{adv}$ is found in the same way as the first technique. However, once the deposition is completed, the substrate is subjected to vibration to achieve the state in which the drop has the minimum Gibbs energy, i.e. the most stable equilibrium state\cite{Mittal_2009}. Subsequently, $\theta_{eq}$ is determined (see Figure \ref{sfig:equib_angle}). Further details on this part can be found in a previous work \cite{Demirel_2020_new}. The volume of each drop in both techniques was set to 10 µL as small amounts of drops affect the accuracy of the measurements \cite{Huhtam_ki_2018}. The drops were pumped in and out at a constant pumping speed of 0.1 µL/s. The capillary numbers of a drop spreading over a used electrode typically fall within the range of $\mathcal{O}(10^{-7} - 10^{-8})$.

\textcolor{blue}{Figure 5} shows that the dynamic contact angles and the contact angle hysteresis, i.e. the difference between $\theta_{adv}$ and $\theta_{rec}$. For a new electrode, the hysteresis remains around 20° across all electrolytes. However, it elevates to around 70$\degree$ for a used electrode. This substantial change provides clear evidence that electrolysis experiments induce significant alterations in the electrode surface. Despite the pronounced differences in dynamic contact angles, $\theta_{eq}$ for both new and used electrodes are remarkably similar, as depicted in Figure \ref{sfig:equib_angle}.

Li et al. \cite{Li2015} reported a water contact angle of $\approx 81^\circ$ for flat platinum film electrodes based on sessile droplet experiments. Fernandez et al. \cite{Fernandez_2014} followed the water contact angle of electrolytically growing bubbles and found a value $\approx68^\circ$ for the case where a single bubble detaches and a new bubble forms. Hydrophobic and other surface contaminants are well-known to lead to finite contact angles on platinum, leading to the use of flame or strong acid treatment to obtain perfect wetting behaviour in the case of Wilhelmy plate measurements \cite{Bewig1965, Momsen1990}. In our case, we measure the ‘most stable’ contact angle as an approximation of the equilibrium contact angle for our platinum electrode material and this value may be quite different compared to other Pt-electrodes depending on the nature of surface contaminants or pretreatments.


\newpage
\clearpage

\section{Spreading Bubbles at Various Conditions}
\label{SI_spreading_all}

  \begin{figure}[ht]
		\centering
		\includegraphics[width=1\columnwidth]{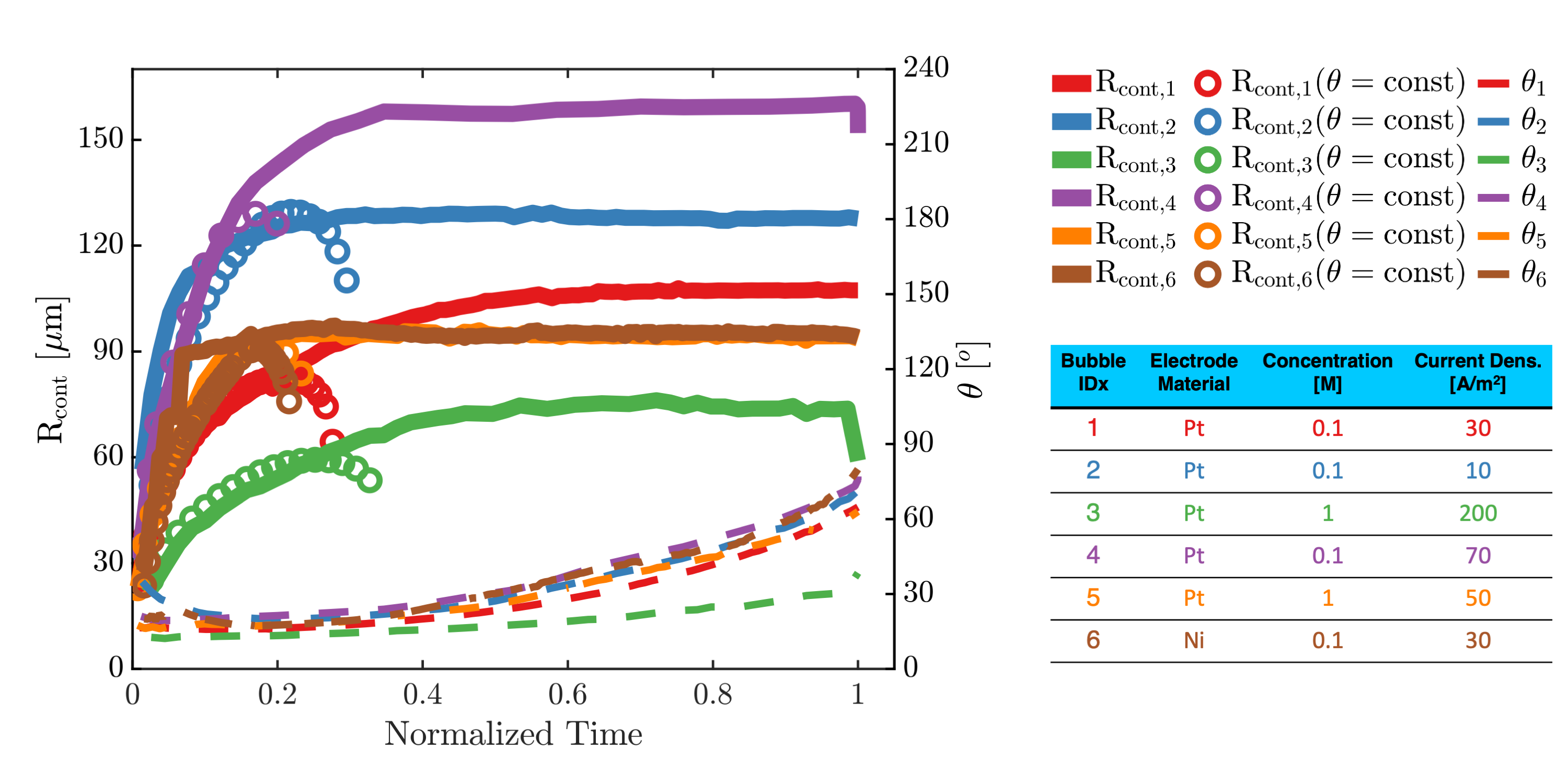}
		\caption{Contact radius ($R_{cont}$) and contact angle ($\theta$) change in time for different bubbles at various experimental conditions.}
		\label{sfig:Rcont_Theta}
	\end{figure}
\bibliography{SI}